\documentclass[12pt,preprint]{aastex}
\usepackage{natbib}
\pdfoutput=1

\begin{document}

\title{Spatially and Spectrally Resolved Hydrogen Gas within 0.1 AU of T Tauri and
Herbig Ae/Be Stars}

\author{J.A. Eisner\altaffilmark{1},  J. D. Monnier\altaffilmark{2}, J. Woillez\altaffilmark{3},
R.L. Akeson\altaffilmark{4},  R. Millan-Gabet\altaffilmark{4},
J.R. Graham\altaffilmark{5}, 
L. A. Hillenbrand\altaffilmark{6}, J.-U. Pott\altaffilmark{7}, S. Ragland\altaffilmark{3},
\& P. Wizinowich\altaffilmark{3}}

\altaffiltext{1}{Steward Observatory, University of Arizona, Tucson, AZ 85721}
\altaffiltext{2}{Astronomy Department, University of Michigan, Ann Arbor, MI 48109}
\altaffiltext{3}{W.M. Keck Observatory, Kamuela, HI 96743}
\altaffiltext{4}{NASA Exoplanet Science Institute, Caltech, Pasadena, CA 91125}
\altaffiltext{5}{Astronomy Department, University of California Berkeley, CA 94720}
\altaffiltext{6}{Astrophysics Department, California Institute of Technology, Pasadena, CA 91125}
\altaffiltext{7}{Max-Planck-Institut f\"{u}r Astronomie, K\"{o}nigstuhl 17, D-69117 Heidelberg, Germany}


\begin{abstract}
We present near-infrared observations of T Tauri and Herbig Ae/Be stars with a
spatial resolution of a few milli-arcseconds and a spectral 
resolution of $\sim 2000$.  Our observations spatially resolve gas and dust
in the inner regions of protoplanetary disks, and spectrally resolve broad-linewidth
emission from the Br$\gamma$ transition of hydrogen gas.  We
use the technique of spectro-astrometry to determine centroids of different
velocity components of this gaseous emission at a precision orders of magnitude
better than the angular resolution. In all sources, we find the
gaseous emission to be more compact than or distributed on similar spatial scales to the
dust emission.  We attempt to fit the data with models including both dust and
Br$\gamma$--emitting gas, and we consider both disk and infall/outflow 
morphologies for the gaseous matter.  In most cases where we can distinguish
between these two models, the data show a preference for infall/outflow models.
In all cases, our data appear consistent with the presence of some gas at stellocentric
radii of $\sim 0.01$ AU.  Our findings support the hypothesis that Br$\gamma$ emission
generally traces magnetospherically driven accretion and/or outflows
in young star/disk systems.
\end{abstract}

\section{Introduction \label{sec:intro}}
Protoplanetary disks play an integral part in the formation of both stars and planets.
Disks provide a reservoir from which stars and planets accrete material, and
a knowledge of the structure of inner regions of disks is needed to understand
the star/disk interface as well as planet formation in disk `terrestrial' regions.
 
Near-infrared interferometry enables spatially resolved observations
of sub-AU-sized regions of protoplanetary disks in nearby star-forming
regions \citep[see][or Dullemond \& Monnier 2010 for recent
reviews]{MILLAN-GABET+07}.  These observations have enabled direct
constraints on the distribution and temperature of dust in terrestrial
planet forming regions, and have
more recently begun to probe gaseous emission as well 
\citep{EISNER+07a,EISNER07,MALBET+07,TATULLI+07,TANNIRKULAM+08,TATULLI+08,ISELLA+08,
KRAUS+08,EISNER+09}.  Spatially resolved observations remove ambiguities
inherent in previous modeling of spatially unresolved spectral energy distributions
\citep[e.g.,][]{BBB88,HILLENBRAND+92} or gaseous emission lines
\citep[e.g.,][]{EDWARDS+94,NCT96}, and allow critical tests of these models.

Spectrally and spatially resolved observations of hydrogen gas have the potential to
constrain how inner disk material accretes onto the central star or
leaves the system (carrying away angular momentum) in outflows.
Because hydrogen in accretion flows or in the 
innermost regions of outflows can be ionized, it may emit 
via a number of electronic transitions as recently recombined atoms
cascade down to the ground state.  
The Br$\gamma$ transition, from the $n=7\rightarrow 4$ electronic states, produces a 
spectral line at 2.1662 $\mu$m, and can be observed with infrared interferometers. 
This line has been shown to be strongly correlated with accretion onto
young stars \citep{MHC98}. While Balmer series hydrogen
lines often show P Cygni profiles associated with winds \citep[or a
combination of winds and infall; e.g.,][]{KHS06}, Br$\gamma$
line profiles are often more consistent with infall kinematics \citep[e.g.,][]{NCT96}.

Accretion is thought to occur by magnetospheric
accretion in low-mass stars \citep[e.g.,][]{KONIGL91}.
Viscous accretion of gas brings material through the disk
to the magnetospheric radius where stellar magnetic fields exert outward pressure
to balance the inward pressure of accretion; gas is then funneled along magnetic field
lines onto high-latitude regions of the star.  The interaction of the stellar magnetic
field and the disk may also lead to the launching of outflows near this 
magnetospheric radius \citep[e.g.,][]{SHU+94} or from stellocentric
radii of an AU or more \citep[e.g.,][]{KP00}.  An alternative to the
magnetospheric accretion picture 
\citep[that may operate in higher-mass stars; e.g.,][]{EISNER+04}
is disk/boundary layer accretion. Matter is accreted viscously through a disk all the way to the star, 
at which point a shock forms due to the large difference in velocities 
between the Keplerian disk and rotating stellar surface \citep[e.g.,][]{LP74}.

Previous, spatially resolved observations of protoplanetary disks
generally found Br$\gamma$ emission to be more compactly distributed than
continuum emission \citep{EISNER07,KRAUS+08,EISNER+09}, although more 
extended distributions were seen in a few cases \citep{MALBET+07,TATULLI+07}.
For the subset of observations where the Br$\gamma$ line was spectrally
resolved, \citet{KRAUS+08} claimed that only one
object appears compatible with a model where the
Br$\gamma$ emission arises in an infalling accretion flow.  For the other four objects in
their sample,  \citet{KRAUS+08} suggested that this emission may trace
extended disk winds \citep[as described in, e.g.,][]{KP00}.  However,
this sample was limited to a few bright A and B stars, which may not be representative of
most young stars.

Here we use the Keck Interferometer (KI) to spatially and spectrally resolve gas
within 1 AU of a sample of fifteen young stars spanning a mass range
from $\sim 0.5$--10 M$_{\odot}$.
These observations expand the previous sample by a factor of 3 in number and by an order of
magnitude in mass range.  We determine the spatial distribution and velocity structure of the 
Br$\gamma$-emitting gas for this sample.  We investigate how these properties depend
on stellar mass or accretion rate, and compare our findings to models
of accretion and outflow.

%In several cases,
%we determine the relative distributions of different gaseous species, 
%including hydrogen, CO, and H$_2$O.

\section{Observations and Data Reduction \label{sec:odr}}

\subsection{Sample \label{sec:sample}}
We selected a sample of young stars (Table \ref{tab:sample})
known to be surrounded by protoplanetary
disks, all of which have been observed previously at near-IR
wavelengths with long-baseline interferometers 
\citep{MST01,EISNER+04,EISNER+05,EISNER+07c,COLAVITA+03,MONNIER+05,
AKESON+05b,AKESON+05}.  All targets have been previously
spatially resolved in the near-IR.  

Our sample (Table \ref{tab:sample})
includes seven T Tauri stars, pre-main-sequence analogs of solar-type
stars like our own sun; five Herbig Ae/Be stars, 2--10 M$_{\odot}$ 
pre-main-sequence stars; and three stars (AS 353, V1057 Cyg, and V1331 Cyg)
with heavily veiled stellar
photospheres whose spectral types are uncertain.  
Our experimental setup
imposes limiting magnitudes of $K \sim 7$ at near-IR wavelengths and 
$V \sim 12$ at optical wavelengths.
We also require that sources be at zenith angles of less than 
$\sim 50^{\circ}$, which excludes from our sample
any sources with $\delta \la -35^{\circ}$. 

The sample was selected to satisfy these criteria, and includes most
T Tauri stars that could be observed, focusing on those that are
known to emit Br$\gamma$ and/or CO overtone emission
\citep[e.g.,][]{FE01,CARR89,NCT96,NAJITA+06}.  
Two sources in our sample are actually fainter than the limiting
system magnitudes at $K$-band: AS 353 and V1331 Cyg.  We included
these because of their strong, previously observed CO overtone emission
\citep[e.g.,][]{CARR89}.  However, as
we discuss below, we were unable to obtain high quality measurements
for these fainter objects.  Finally, we included several brighter Herbig 
Ae/Be stars that expanded the stellar mass range of the sample.

\subsection{Experimental Setup \label{sec:setup}}
KI is a fringe-tracking long baseline near-IR Michelson
interferometer combining light from the two 10-m Keck apertures 
\citep{CW03,COLAVITA+03}.  Each of the 10-m apertures is equipped with
a natural guide star adaptive optics (NGS-AO) system that corrects phase errors caused by
atmospheric turbulence across each telescope pupil, and thereby maintains 
spatial coherence of the light from the source across each aperture.  The
NGS-AO systems require sources with $R$ magnitudes brighter than $\sim 12$.
Optical beam-trains transport the light from each Keck aperture down into
a tunnel connecting the two Kecks and to a set of beam combination optics.

We used the ``self phase referencing''(SPR)  mode of KI, implemented as
part of the ASTrometric and phase-Referenced Astronomy (ASTRA) program
(see Woillez et al. 2010).  
%ASTRA is a program of
%upgrades to KI, funded by an NSF MRI grant.  
%SPR is a stepping stone to
%dual-field phase referencing, which will extend the sensitivity of KI
%by several orders of magnitude in cases where a bright, nearby star
%can be used as a phase reference.  Ultimately, ASTRA will deliver
%an astrometric capability with a projected measurement precision
%of $\sim 30$--50 $\mu$as.  
%SPR is the first major component
%of the ASTRA program that has been commissioned.
The SPR mode introduces a split in
the beams from each aperture, immediately before the
fast delay line optics.  55\% of the light is passed down the ``primary'' channel,
which consists of the normal KI optics used in the standard $V^2$ mode.
Of the remaining 45\%, 20\% of the light is split off to the ``secondary'' fast delay lines
(which are generally used as part of the KI Nuller)\footnote{This beamsplitting optic was designed
to split $K$ and $N$ band light for the Nuller, and is not perfect as a $K$-band beamsplitter.},
and ultimately sent to a second beam-combining table and detector.
Each of the detectors consists of a HAWAII array.

For both primary and secondary sides, interferometric fringes are measured by modulating 
the relative delay of the two input beams and then measuring the modulated intensity level 
of the combined beams during four ``ABCD'' detector reads \citep{COLAVITA99}.
Each read has an integration time of 2 ms.
The measured intensities in these reads are used to determine
atmosphere-induced fringe motions, and a servo loop removes these motions
to keep the fringes centered near zero phase.

For the secondary side,
a servo loop uses the phase information measured on the primary side to
stabilize the atmospheric phase motions, and we can thus use longer modulation periods
(and hence integration times).  We use integration times between
0.5 and 2 s, approximately 1,000 times longer than possible with the uncorrected
primary side.  While integrations longer than 2 s are possible, observations at wavelengths
longer than 2 $\mu$m become background-limited in this regime.

In front of the secondary detector is a grism.
This grism consists of a prism made of S-FTM16 glass, with refractive index$=$1.56 and
an apex angle of 36.8$^{\circ}$; and
an epoxy grating with 150 grooves per mm and a blaze angle of 36.8$^{\circ}$.
Used in first-order, the grism passes the entire $K$-band
with a dispersion of $\lambda/\Delta \lambda \approx 2000$.  
This spectral resolution is confirmed with measurements of a neon lamp
spectrum.  Note, however, that the lines are not fully Nyquist sampled with
our detector; spectra are Nyquist sampled at a resolution of $\sim
1000$.   Neon lamp spectra and/or Fourier Transform Spectroscopy are
also used to determine the wavelength scale for each night of observed data.
While the entire $K$-band
falls on the detector, vignetting in the camera leads to lower throughput toward the band 
edges.  The effective bandpass of our observations is approximately 2.05 to 2.35 $\mu$m.

In this paper we focus on Br$\gamma$ emission, which fills only
a small portion of the $K$-band.  Our measurements actually cover
other interesting spectral regions that include significant opacity
from H$_2$O and CO transitions.  We defer discussion of these
spectral regions to a later paper.

\subsection{Observations \label{sec:obs}}
We obtained Keck Interferometer (KI) observations of our sample on 
UT 2008 April 25, 2008 November 17, 2008 November 18, and 2009 July 15
(see Table \ref{tab:obs}).
The first of these nights was actually the commissioning night of the SPR mode, 
and so we observed a number of unresolved calibrator
stars and known, strong Br$\gamma$ emitters, to use for system characterization.  The analysis of these initial
data is presented in separate papers (Woillez et al. 2010; Pott et al. 2010).  When observing
our sample, targets were interleaved with calibrators every 10--15 minutes.

\subsection{$V^2$ Calibration \label{sec:v2cal}}

We measured squared visibilities ($V^2$) for our targets and calibrator
stars in each of the 330 spectral channels across the $K$-band provided by the grism. 
The calibrator stars are main
sequence stars, with known parallaxes,
whose $K$ magnitudes are within 0.5 mags of the target $K$ magnitudes
(Table \ref{tab:sample}).
The system visibility (i.e., the point source response of the interferometer)
was measured using observations of these calibrators, whose angular sizes
were estimated by fitting blackbodies 
%(with the temperature constrained by the
%measured spectral type) 
to literature photometry.  These size estimates are not crucial since the
calibrators are unresolved (i.e., their angular sizes are much smaller
than the interferometric fringe spacing) in almost all cases.  
HD 163955, a calibrator
for MWC 275, is mildly resolved; we account for this when computing the
system visibility.

We calculated the system visibility appropriate to each target scan
by weighting the calibrator data by the internal scatter and the temporal and 
angular proximity to the target data \citep{BODEN+98}.  For comparison, we 
also computed the straight average of the $V^2$ for all calibrators used for a 
given source, and the system visibility for the calibrator observations 
closest in time.  These methods all produce results consistent within the 
measurement uncertainties.  We adopt the first method in the analysis that 
follows.

Source and calibrator data were corrected for 
standard detection biases as described by \citet{COLAVITA99} and averaged into 
5 s blocks. 
%We accounted for a known bias related to flux level
%by applying an empirically determined 
%correction\footnote{http://msc.caltech.edu/software/KISupport/dataMemos/fluxbias.pdf}.  
Calibrated $V^2$ were then computed by dividing the average
measured $V^2$ over 130 s scans (consisting of 5 s sub-blocks) for
targets by the average system visibility.  
Uncertainties are given by the quadrature 
addition of the internal scatter in the target data
and the uncertainty in the system visibility. 
We average together all of the calibrated data for a given source 
to produce a single measurement of 
$V^2$ in each spectral channel.  The observations of our targets typically
spanned $\la 1$ hour, and the averaging therefore has a 
small effect on the uv coverage.  
%Previous observations showed that
%this technique produces channel-to-channel uncertainties of a few percent
%\citep{EISNER+07b,EISNER07}.  

As seen in previous observations with a lower-dispersion grism at KI
\citep{EISNER+07b}, we find channel-to-channel
uncertainties of a few percent or less in our data.  
Here we estimate these uncertainties by computing the standard
deviation of $V^2$ measured in a spectral region spanning
2.2 to 2.25 $\mu$m.  This region typically
has good signal-to-noise, and does not contain signal
from Br$\gamma$ emission or absorption. 

The normalization of $V^2$ versus wavelength 
(i.e., the average value of $V^2$ across the band) has an additional 
uncertainty of $\sim 3\%$.  We ignore this in our analysis since
it does not affect the relative measurements of various channels.

\subsection{Differential Phase Calibration \label{sec:dpcal}}

The ``ABCD'' reads are used to calculate
the phase of interference fringes recorded in both the primary channel
(as described in \S \ref{sec:setup}) and in the secondary channel.
Infrared interferometers generally do not
measure phase information that is intrinsic to the target, because
rapid atmospheric phase distortions  scramble the phase of the ``pristine''
wavefronts emitted by the source \citep[e.g.,][]{MONNIER07}.
However, with our SPR observations the atmospheric phase
distortions are measured by the primary fringe tracker, and
these phase motions are subtracted from the secondary side.
Furthermore, for the dispersed fringes on the secondary side,
any residual atmospheric phase motions would affect the
fringes in each channel (approximately) the same way,
and so the differential phase ($\Delta \phi$) across the observed spectrum
is intrinsic to the source.

Raw phases are measured in the same ``ABCD'' reads used
to determine $V^2$.  These phases are then de-rotated so 
that the average phase of all channels is zero.  Next,
the phases versus wavelength are unwrapped to 
eliminate any 180$^{\circ}$ jumps.  After computing weighted average
differential phases for each target and calibrator scan,
we determine a ``system differential phase'' using similar weighting
employed above to calculate the system visibility.

The system differential phase is subtracted from the target
differential phase.  Since targets
and calibrators are observed at similar airmasses, this
calibration procedure removes most atmospheric and
instrumental refraction effects.  Finally, we remove any
residual slope in the differential phase spectrum, since
we can not distinguish instrumental slopes from those
intrinsic to the target signal.  Since we are focused
on a small spectral region around the Br$\gamma$ feature,
we are largely insensitive to errors in these calibrations.

We estimate the uncertainty on the differential phase
measurements by taking the standard deviation of 
measurements in a spectral region between 2.2 and 2.25 $\mu$m
(as we did for $V^2$ above).  The errors
derived from the data are typically $\la 1^{\circ}$, although the faintest
objects in our sample exhibit somewhat larger uncertainties.

\subsection{Flux Calibration \label{sec:fluxcal}}

We used the count rates in each channel observed during 
``foreground integrations'' \citep{COLAVITA99} to recover crude
spectra for our targets.  These spectra are measured when no fringes
are present, and include all flux measured
within the $\sim 50$ mas diameter of the instrumental field of view.
We divided the measured flux versus wavelength
for our targets by the observed fluxes from the calibrator stars, using
calibrator scans nearest in time to given target scans, and then multiplied
the results by template spectra suitable for the spectral types of
the calibrators.  

We used Nextgen stellar atmosphere models \citep{HAB99} as templates.
These model spectra are computed at a resolution of 2 \AA (more than
five times finer than the resolution of our data) for stars with 
effective temperatures up to 10,000 K.  They 
are thus suitable templates to be used
with our calibrator stars (see Table \ref{tab:sample}).
%We note that while we have used Kurucz models in the past 
%\citep[e.g.,][]{EISNER+09}, the spectral resolution in these
%models is too low to be used for calibrating the current dataset.

Tests of our calibration procedure for
main sequence stars of known spectral type, calibrated using other
calibrator stars, indicate channel-to-channel uncertainties  of a few
percent.  Similar uncertainties are found by calculating the standard deviation
in a spectral region adjacent to Br$\gamma$; we adopt these
uncertainties for the analysis presented below.  While
we also see evidence for uncertainties in the spectral slope 
across the  $K$-band (the slope varies from one observation to the
next), we ignore these here since they do not affect the narrow-band data 
in the Br$\gamma$ spectral region.

\subsection{Separating Stellar and Circumstellar Components \label{sec:ratios}}
The fluxes, squared visibilities, and differential phases described
above contain contributions from both circumstellar material and
central stars.  Since our interest here is in the
circumstellar matter, we remove the stellar component of the
measurements before proceeding.

Decomposition of the spectrum into stellar and circumstellar components
is straightforward as long as the ratio of circumstellar-to-stellar
flux is known at each observed wavelength.  
%Because the central stars are unresolved, we know that $V^2_{\ast}=1$, 
%and so the circumstellar-to-stellar flux ratio is sufficient for removing the
%stellar contribution to the measured $V^2$ as well.  Finally, 
%if we assume the the star sits at the zero phase position--as it will
%if the circumstellar distribution is symmetric and centered on the star--then
%the circumstellar-to-stellar flux ratio can
%be used to remove the stellar contribution to the differential phase signal.
We estimate the circumstellar-to-stellar flux ratio at each observed
wavelength following \citet{EISNER+09}.  We use
stellar radii, effective temperatures, distances, and extinctions (assuming
the reddening law of \citet{ST91}) from the literature \citep{HG97,MMD98,WG01,
MUZEROLLE+03,EISNER+04,EISNER+05,MONNIER+05,MONNIER+06,BSC07}
to fit the stellar photosphere.  The Nextgen models described in \S \ref{sec:fluxcal} are
then used to determine the stellar fluxes at each of our observed wavelengths.
These models are suitable for all of our stars except MWC 1080, whose 
effective temperature is substantially higher than the 10,000 K maximum
of the Nextgen models.  Such hot stars typically have shallower Br$\gamma$
absorption than cooler A stars, since the H ionization fraction in the stellar photosphere is higher.
Thus, the circumstellar Br$\gamma$ spectrum may underestimate the true
line-to-continuum ratio.

For AS 353A, V1331 Cyg, and V1057 Cyg, whose stellar photospheres are
effectively invisible, we can not reliably estimate the circumstellar-to-stellar flux ratios.
For these objects we assume that 100\% of the $K$-band flux arises from the
circumstellar environments.  This assumption is consistent with previous 
spectroscopic observations of these objects \citep{ESB90,HJ83,EISNER+07c,HHC04}.

To estimate the circumstellar components of the $V^2$ and $\Delta
\phi$, we consider the contribution of the star to the {\it complex} visibilities:
\begin{equation}
V_{\rm meas} e^{i \Delta \phi_{\rm meas}} = \frac{F_{\ast} V_{\ast}  e^{i
  \Delta \phi_{\ast}} +   F_{\rm disk} V_{\rm disk}  e^{i \Delta
  \phi_{\rm disk}}}{F_{\ast}+F_{\rm disk}} .
\end{equation}
Here, $V_{\rm meas}=\sqrt{V^2_{\rm meas}}$, $F_{\ast}$ and $F_{\rm
  disk}$ are the stellar and circumstellar fluxes (determined above),
and $V_{\ast}$ and $V_{\rm disk}$ are the stellar and circumstellar
components contributing to the measured visibilities.
Assuming the central star is unresolved and at the
photo-center of the continuum emission, this can be simplified to
\begin{equation}
V_{\rm meas} e^{i \Delta \phi_{\rm meas}}  =\frac{F_{\ast} +   F_{\rm disk} V_{\rm disk}  e^{i \Delta
  \phi_{\rm disk}}}{F_{\ast}+F_{\rm disk}}.
\label{eq:vc_disk}
\end{equation}
The real and imaginary parts of Equation \ref{eq:vc_disk} provide two
equations, which can be solved for the two
unknown quantities $V_{\rm disk}$ and $\Delta \phi_{\rm disk}$:
\begin{equation}
\Delta \phi_{\rm disk} = \tan^{-1} \left[ \frac{V_{\rm meas} \sin
    (\Delta \phi_{\rm meas}) (F_{\ast} + F_{\rm disk})}{V_{\rm meas}
    \cos (\Delta \phi_{\rm meas}) (F_{\ast}+F_{\rm disk}) - F_{\ast}}\right],
\label{eq:dpcirc}
\end{equation}
\begin{equation}
V_{\rm disk} = V_{\rm meas} \frac{\sin (\Delta \phi_{\rm meas})}{\sin
  (\Delta \phi_{\rm disk})} \left(\frac{F_{\ast}+F_{\rm disk}}{F_{\rm disk}}\right).
\label{eq:vdisk}
\end{equation}

We use Equations \ref{eq:dpcirc} and \ref{eq:vdisk} to compute the
squared visibilities and differential phases corresponding to the
circumstellar matter. However, we note that if the differential phases
are small  ($\ll 1$ rad), we can simplify these equations to somewhat
more intuitive forms using the small angle approximation:
\begin{equation}
V_{\rm disk} = \frac{V_{\rm meas}(1+F_{\rm disk}/F_{\ast})-1}
{F_{\rm disk}/F_{\ast}},
\label{eq:v2disk}
\end{equation}
\begin{equation}
\Delta \phi_{\rm disk} =\frac{ \Delta \phi_{\rm meas} \left(1+V_{\rm
      disk} F_{\rm disk}/F_{\ast}\right)} {V_{\rm disk} F_{\rm disk}/F_{\ast}}.
\label{eq:dpdisk}
\end{equation}
This assumption is equivalent to assuming that the differential phase
is linearly related to the centroid offset, which is true for
unresolved or marginally resolved structures:
\begin{equation}
\Delta \theta = \Delta \phi \frac{\lambda}{2\pi B},
\label{eq:dp}
\end{equation}
where $\lambda$ is the observed wavelength and $B$
is the projected baseline length.  

The uncertainties on $V^2_{\rm disk}$ and $\Delta \phi_{\rm disk}$ are 
computed from the uncertainties in $V^2_{\rm meas}$ or $\Delta
\phi_{\rm meas}$ and an assumed uncertainty of 20\% on the circumstellar-to-stellar
flux ratio.  We fit our models for the circumstellar emission to 
$F_{\rm disk}$, $V^2_{\rm disk}$, and $\Delta \phi_{\rm disk}$  below.

\subsection{General Features of Calibrated Data \label{sec:preresults}}
\begin{figure}[tbhp]
\plotone{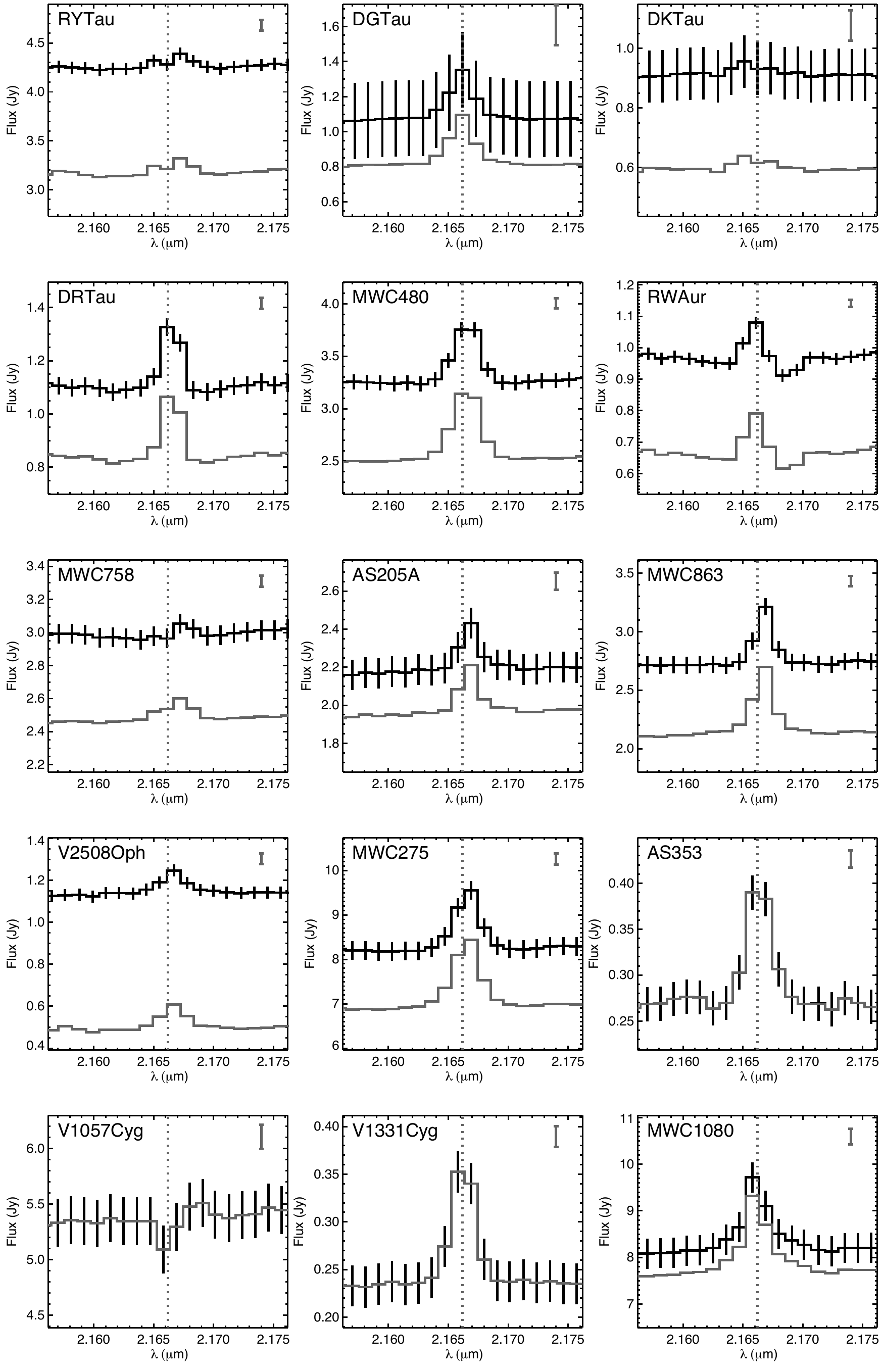}
\caption{Spectra of our sample in the spectral region around Br$\gamma$.  Observed fluxes
are plotted with solid black histograms.  Circumstellar
fluxes, determined using the procedure described in \S \ref{sec:ratios}, are shown with
gray histograms.  For AS 353, V1057 Cyg, and V1331 Cyg, we assume that all of
the near-IR flux arises from the circumstellar environment.  For
clarity of presentation, we have not plotted the error bars associated
with the circumstellar fluxes; the magnitudes of the uncertainties are
indicated in the upper right corners.  Dotted
gray lines indicate the
central (rest) wavelength of the Br$\gamma$ transition, 2.1662 $\mu$m.
\label{fig:spectra}}
\end{figure}

Figure \ref{fig:spectra} shows the fluxes in the Br$\gamma$ spectral region,
calibrated with the procedure outlined in \S \ref{sec:fluxcal}.  Since many of our targets are early-type
stars, which exhibit photospheric Br$\gamma$ absorption, we also
plot in Figure \ref{fig:spectra} spectra from which the stellar
contribution has been removed (\S \ref{sec:ratios}).
All of our targets except one
show evidence of circumstellar Br$\gamma$ emission; in V1057 Cyg,
Br$\gamma$ appears in absorption.
In the analysis presented below we will focus 
on the remaining fourteen objects that exhibit Br$\gamma$
emission.

\begin{figure}[tbhp]
\plotone{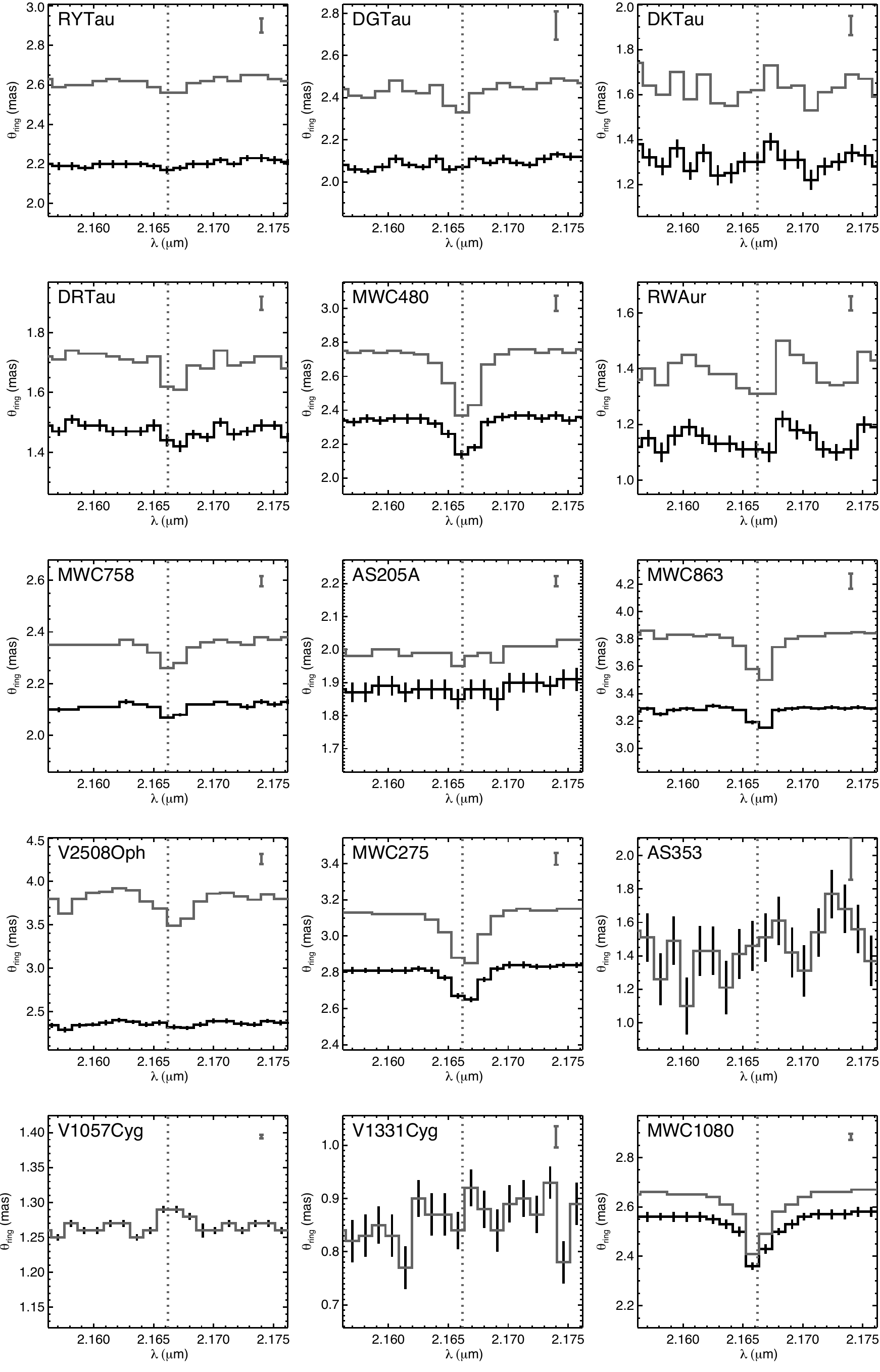}
\caption{Uniform ring angular diameters of our sample, plotted in the spectral region
around Br$\gamma$.  Angular sizes computed directly from the observed $V^2$ are plotted
with black histograms.  Angular sizes of only
the circumstellar emission, determined using the procedure described in \S \ref{sec:ratios}, are 
shown with gray histograms. 
\label{fig:uds}}
\end{figure}

To better illustrate the behavior of $V^2$ versus wavelength, we 
fit the $V^2$ data for each source, in each channel, with 
a simple uniform ring model \citep[e.g.,][]{EISNER+04}.
The results, shown in Figure \ref{fig:uds}, 
give the ``spectral size distribution'' of the
emission, illustrating how the spatial scale of the near-IR
emission depends on wavelength.  Figure \ref{fig:uds}
also shows the estimated angular size of only the circumstellar
emission, calculated as described in \S \ref{sec:ratios}

Figure \ref{fig:uds} shows that the angular diameter of the near-IR
emission changes across the spectral region of Br$\gamma$
emission in most of our target objects.  While more compact sizes
of Br$\gamma$ emission relative to continuum emission
have been previously reported for a number of these objects
\citep{EISNER07,EISNER+09}, here we resolve the size
versus wavelength spectrally across the Br$\gamma$ feature.
This augments the small number of existing measurements
of this kind, since our sample only has an overlap of one
with previous observations at similar spectral resolution
\citep{KRAUS+08}.  

In all of our targets, the Br$\gamma$ emission
appears either more compactly distributed than the
surrounding continuum emission, or distributed on
similar spatial scales.  Using Equations
\ref{eq:dpcirc} and \ref{eq:vdisk}, but with the line-to-continuum
flux ratio in each spectral channel in lieu of the
circumstellar-to-stellar flux ratios, we can compute the $V^2$ for the
Br$\gamma$-emitting gas.  Fitting this $V^2$ with a uniform ring
model, we can derive simple size estimates. In particular,
we fit a uniform ring model to the peak channel; since this is the
lowest velocity material, it should also be the most extended and
should thus trace approximately the full extent of the
Br$\gamma$--emitting gas.  

In Table
\ref{tab:brgsizes}, we compare the derived uniform ring sizes for the
Br$\gamma$ emitting region with those of the continuum.
Table \ref{tab:brgsizes} confirms the suggestion, based on Figure
\ref{fig:uds}, that the Br$\gamma$ emission is more compact than, or
distributed on similar scales to, the continuum emission for all
objects in our sample.  Furthermore, we see that in a number of cases,
the inferred size of the Br$\gamma$ emission is $\la 0.1$ mas,
corresponding to radii $\la 0.01$ AU for our targets.

\begin{figure}[tbhp]
\plotone{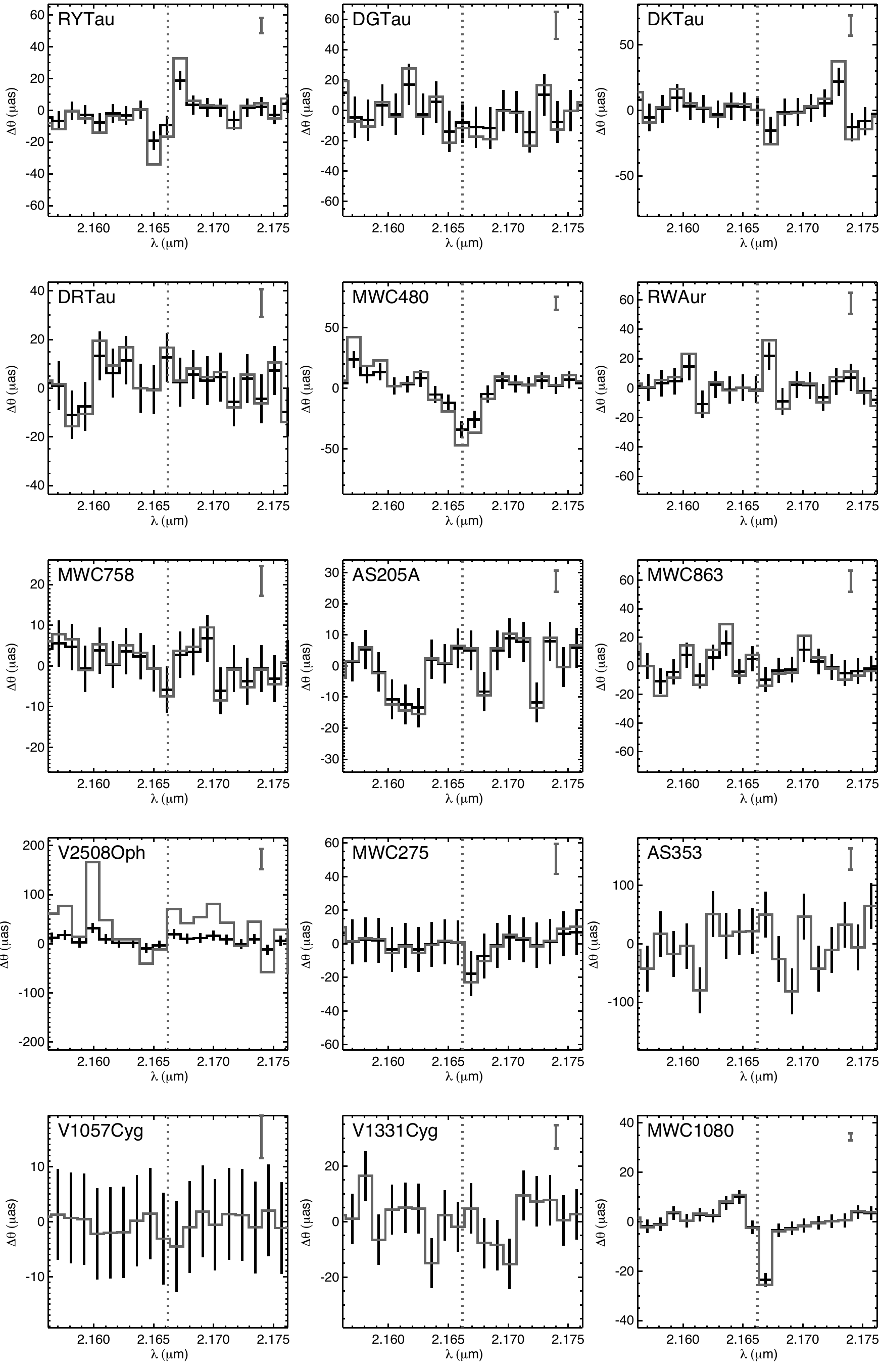}
\caption{Centroid offsets of our sample, plotted in the spectral region
around Br$\gamma$.  Black histograms show the offsets derived for the observed data, and gray histograms
show the centroid offsets for the circumstellar component of the emission.
\label{fig:cents}}
\end{figure}

To facilitate interpretation of the differential phase data,
we convert the $\Delta \phi$ values into centroid offsets
(Equation \ref{eq:dp}).
Since KI consists
of a single baseline, and we observe our targets
with a limited $uv$ range, the derived centroid offsets
are essentially projections onto a single position angle
on the sky for a given object.  The centroid offsets
versus wavelength for our sample are plotted in Figure
\ref{fig:cents}.

In most cases, no clear signals are seen in the differential phase
data above the uncertainty level. However, a few objects display
deviations from the mean phase at the wavelength of the Br$\gamma$
transition.  Objects with clear signals are generally the brightest
objects in our sample (RY Tau, MWC 480, MWC 275, and MWC 1080).  The
differential phase data is noisier and less reliable for fainter
sources.  For example, while AS 205 exhibits features in the
differential phases (not at the Br$\gamma$ wavelength) that appear
significant, these may be spurious signals for this fainter object.

\section{Modeling \label{sec:mod}}

In this section we use our flux, $V^2$,  and differential phase
measurements, after removal of the stellar components (\S
\ref{sec:ratios}) to constrain the distribution of dust and gas around
our sample stars.  In \S \ref{sec:diskmods}, we consider
a Keplerian disk model with a rotating, gaseous disk extending from
some inner radius out to the dust sublimation radius, where a ring of
continuum emission resides.  In \S \ref{sec:infall}, we replace the
rotating, gaseous disk with a compact, bipolar infall/outflow structure.  

In these models we make the simple assumption that continuum emission
is confined to a ring whose annular width is 20\% of its inner radius.
%\citep[consistent with previous modeling; e.g.,][]{EISNER+04}.   
This assumption makes sense if the continuum
emission traces dust, since most of the emission will come from radii
near the sublimation point of silicate dust \citep{DDN01,IN05}.
However, previous work has shown that inner
disk gas or highly refractory dust, on smaller scales than the bulk of
the dust emission, also produces continuum emission
\citep[e.g.,][]{TANNIRKULAM+08,EISNER+09,BENISTY+10}.
This compact continuum emission can lead to underestimated sizes for
the (bulk) dust distribution.  However, since we are focused on the line
emission, we are not concerned with such potential errors.  The
assumption that continuum emission lies in a single ring should not affect
inferred constraints on the Br$\gamma$ emission morphology.

The properties of the continuum emission can be constrained directly
from the observations of spectral regions free of Br$\gamma$ emission.
The inner ring radius is determined  directly from a fit of a uniform
ring model to $V^2$ data in spectral regions adjacent to those where Br$\gamma$ 
emission is observed.  We determine the ring radius
for all position angles and inclinations considered for
our gaseous disk model (below).  With $R_{\rm ring}$ determined from the
$V^2$ data, the temperature of the ring is set so that 
the continuum flux level matches the observed continuum fluxes.
Neither $R_{\rm ring}$ nor $T_{\rm ring}$ are free parameters in the
models discussed below.

\subsection{Keplerian Disks \label{sec:diskmods}}
We begin with a model that assumes all of the observed emission,
including both continuum and Br$\gamma$, lies in a common
disk plane.   The gaseous disk extends from $R_{\rm in}$ to  $R_{\rm out}=R_{\rm ring}$.
While gaseous emission may exist at stellocentric radii larger than
$R_{\rm ring}$, we make the simple assumption that any such emission
is hidden by the optically thick dust disk (we ignore any potential emission
from hot gas in outer disk surface layers).
Both the disk and the ring of continuum emission have a
common  inclination,
$i$, and position angle, $PA$, that are  free parameters.

The brightness profile of the gaseous disk is parameterized with 
a power-law, 
\begin{equation}
B_{\rm disk} (R) = B_{\rm in} \left(\frac{R}{R_{\rm in}}\right)^{-\alpha}. 
\end{equation}
The value of $\alpha$ is difficult to determine analytically since
it depends on the temperature profile and surface density profile
of the disk.  We therefore leave 
$\alpha$ as a free parameter in our modeling.  

The normalization of the brightness profile is chosen so that the
resulting spectrum has a specified line-to-continuum ratio, $L/C$.  
$L/C$ is defined as the ratio of the total flux of the gaseous
emission, integrated over space and velocity, to the total flux of the
continuum component.

We assume the gas to be in Keplerian rotation, with a radial velocity profile,
\begin{equation}
v_{\rm obs}(R) = \sqrt{\frac{G M_{\ast}}{R}} \cos (PA) \sin (i).
\end{equation}
Here, $M_{\ast}$ is the stellar mass, $PA$ is the disk position angle,
and $i$ is the disk inclination.  For simplicity,
we do not enter exact values of $M_{\ast}$ for
each source into the model (these are not determined
to high accuracy for most objects).  Rather, we assume
a stellar mass of 1 M$_{\odot}$ for the T Tauri stars in our sample, 
3 M$_{\odot}$ for the Herbig Ae stars, and 10 M$_{\odot}$ for
the Herbig Be star MWC 1080.

Using the power-law brightness profile and (assumed) Keplerian
velocity profile, we generate channel maps of the model.  Each modeled channel
is centered on the central wavelength of a channel in our data.
Channel maps are generated
and saved for a grid of parameter values.

To fit this model to the data for a given source, we must first convert
the model channel maps into spectra, squared visibilities, and differential phases.
The model spectra are computed by summing the flux in each channel map.
We then calculate a discrete Fourier Transform of each channel map to determine
the complex visibilities we would ``observe'':
\begin{equation}
V_{\rm model} (u,v) = \frac{\sum_{x} \sum_{y} A(x,y) e^{-2 \pi i (u x + v y)/\lambda}} 
 {\sum_{ x} \sum_{y} A(x,y)} .
\end{equation}
Here, $A(x,y)$ is the brightness distribution of a given channel map,
$u$ and $v$ are the projected east-west and north-south baseline lengths, and the
sums are taken over the spatial dimensions $x$ and $y$.
The model $V^2$ and $\Delta \phi$ are the squared amplitude 
and phase of these complex visibilities:
\begin{equation}
V^2_{\rm model} = {\rm Real}\{V_{\rm model}(u,v)\}^2 + 
 {\rm Imaginary}\{V_{\rm model}(u,v)\}^2,
\end{equation}
\begin{equation}
\Delta \phi_{\rm model} = \tan^{-1} \left(\frac{ {\rm Imaginary}\{V_{\rm model}(u,v)\}}
{ {\rm Real}\{V_{\rm model}(u,v)\}}\right).
\end{equation}

\begin{figure}[tbp]
\plotone{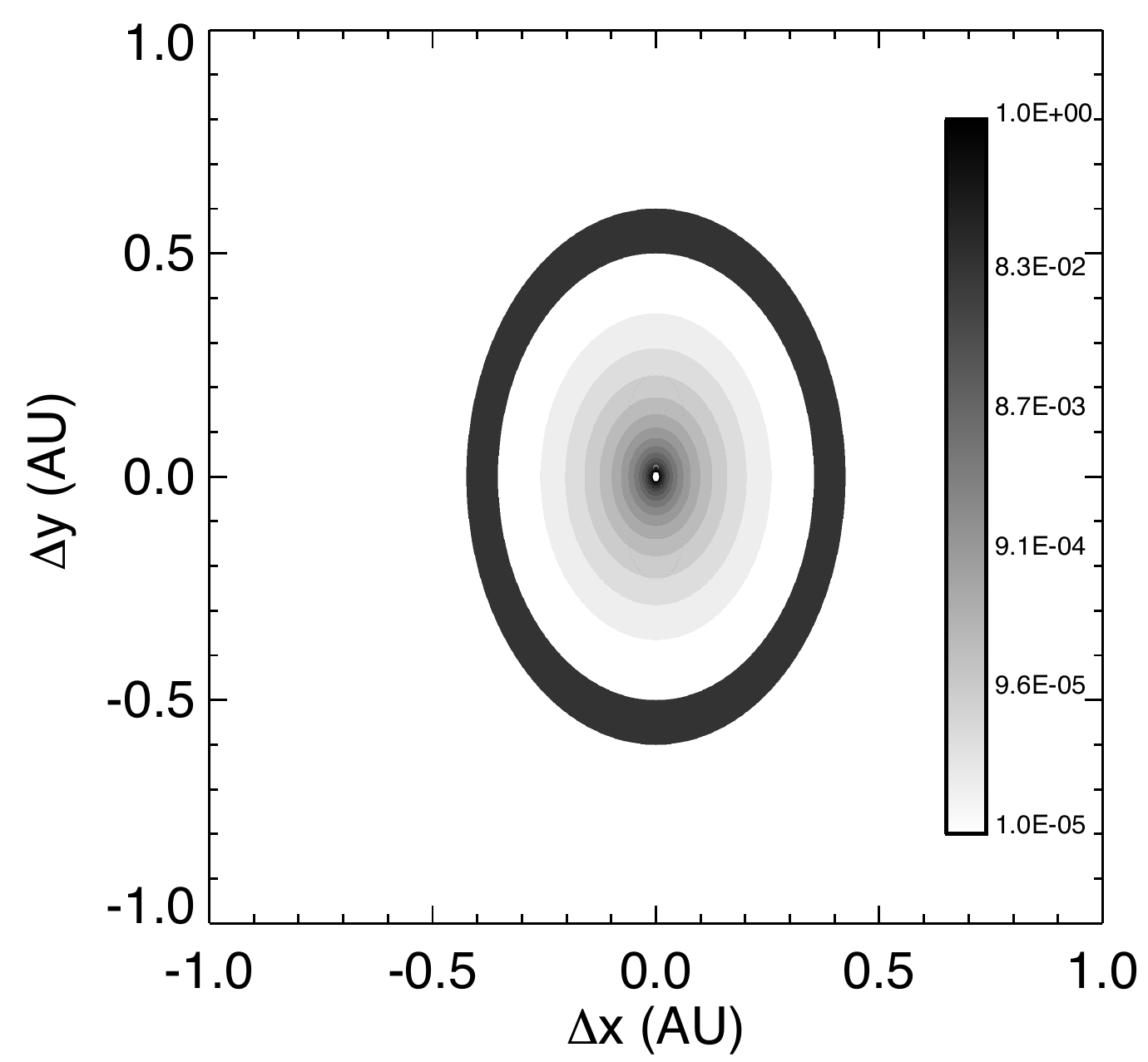}
\caption{Synthetic, velocity-integrated image, computed
for the disk model described in \S \ref{sec:diskmods} assuming
$R_{\rm ring}=0.5$ AU, $R_{\rm in}=0.01$ AU, PA=0, $i=45^{\circ}$, 
$\alpha=3$, a line-to-continuum ratio of 0.5, and $M_{\ast}=3$ M$_{\odot}$. 
The image has been normalized, and is shown with a logarithmic
stretch.  Note that the star is not included in this synthetic image.
\label{fig:diskimage}}
\end{figure}

\begin{figure}[tbp]
\plotone{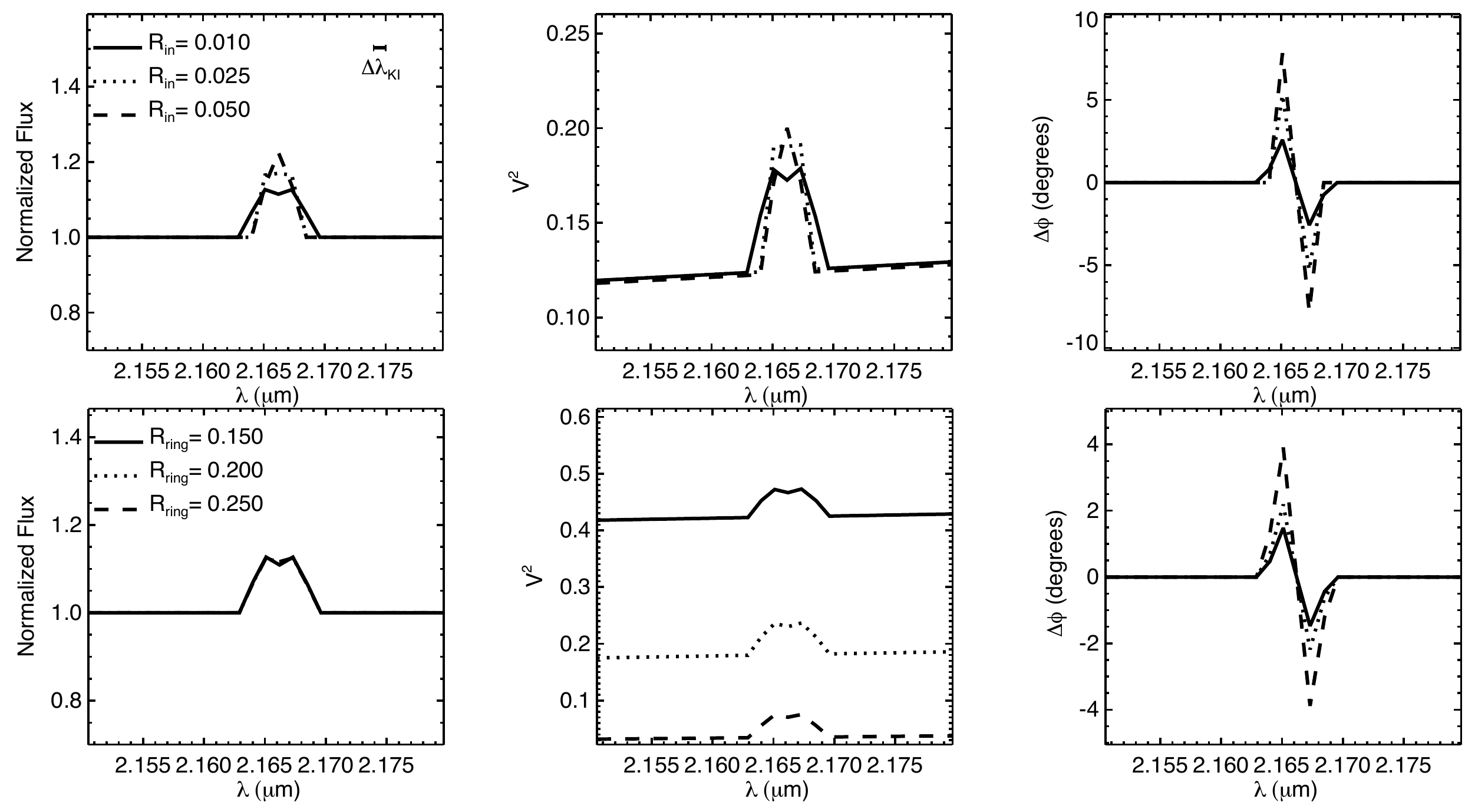}
\caption{Synthetic fluxes, $V^2$, and differential phases ($\Delta \phi$) computed
for the disk model described in \S \ref{sec:diskmods}.  We assume a fiducial model
with $R_{\rm in}=0.01$ AU, PA=0, $i=45^{\circ}$, $\alpha=3$, a
line-to-continuum ratio of 0.5, $d=140$ pc, and $M_{\ast}=3$
M$_{\odot}$.  Here, we vary $R_{\rm in}$ ({\it top}) and $R_{\rm
  ring}$ ({\it bottom}), and illustrate the effects of these
variations on the synthetic data.
\label{fig:disk_geometry}}
\end{figure}

\begin{figure}[tbp]
\plotone{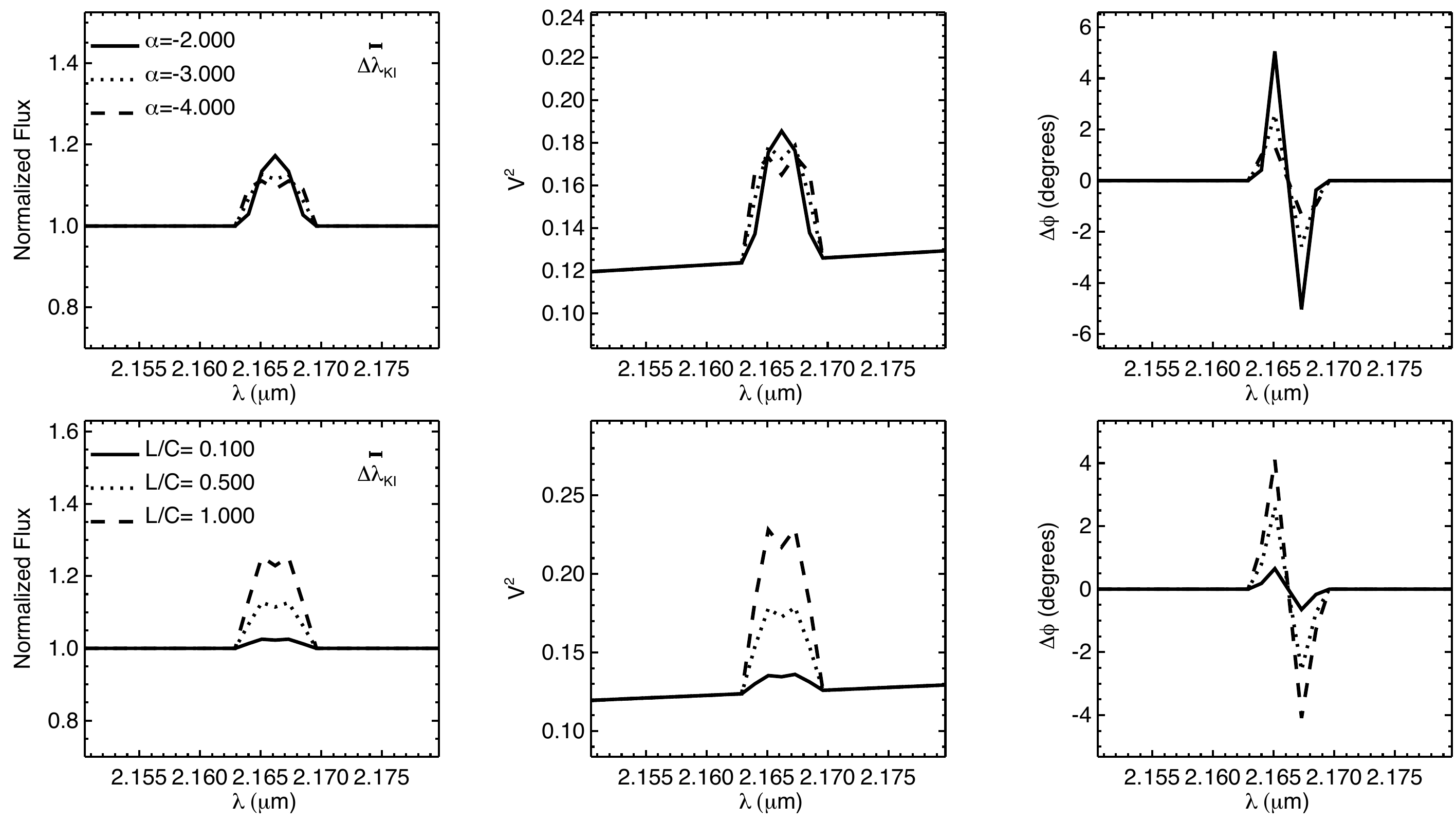}
\caption{Synthetic fluxes, $V^2$, and differential phases ($\Delta \phi$) computed
for the disk model described in \S \ref{sec:diskmods}.  Assuming the fiducial model
described in Figure \ref{fig:disk_geometry}, we vary $\alpha$ ({\it top})
and $L/C$ ({\it bottom}) and illustrate the effects of changing these parameters on the synthetic data.
\label{fig:disk_brightness}}
\end{figure}

\begin{figure}[tbp]
\plotone{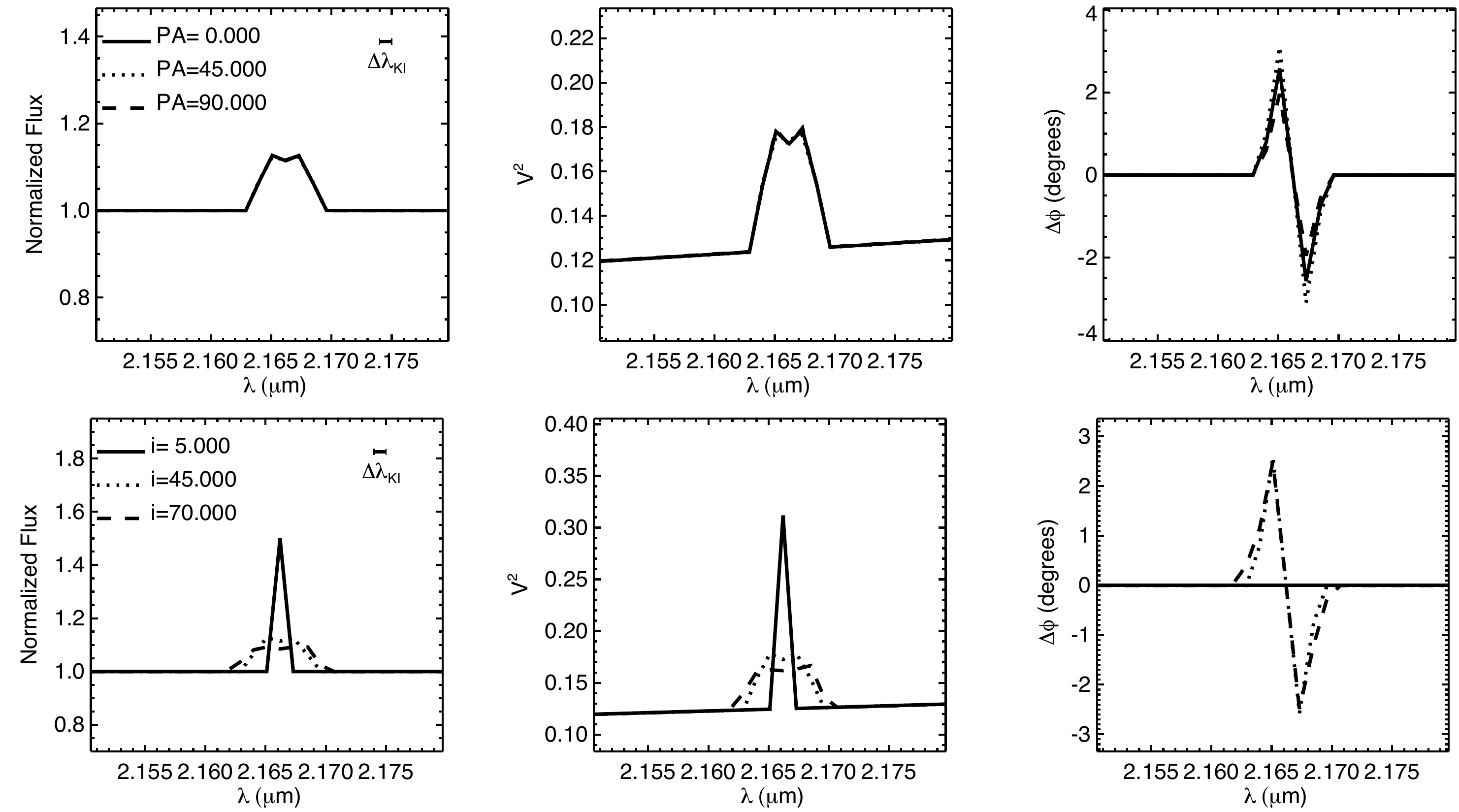}
\caption{Synthetic fluxes, $V^2$, and differential phases ($\Delta \phi$) computed
for the disk model described in \S \ref{sec:diskmods}.  Assuming the fiducial model
described in Figure \ref{fig:disk_geometry}, we vary $PA$ ({\it top})
and $i$ ({\it bottom}) and illustrate the effects of changing these parameters on the synthetic data.
\label{fig:disk_viewing}}
\end{figure}

An image of an example disk model is shown in Figure \ref{fig:diskimage}.
Different disk morphologies can be obtained by adjusting the free
parameters described above.  These parameters can be considered in
three groups: disk geometry, described by $R_{\rm in}$ and $R_{\rm
  ring}$; the disk brightness profile, parameterized by $\alpha$ and
$L/C$; and the viewing geometry, described by $PA$ and $i$.
Before describing the results of fitting these disk models to data, we
start with brief discussion of how model parameters affect synthetic
fluxes, $V^2$, and differential phases.  These effects are illustrated
in Figures \ref{fig:disk_geometry}--\ref{fig:disk_viewing}, and
described in the following sections.

\subsubsection{Disk Geometry}
Here we consider the effects of $R_{\rm in}$ and $R_{\rm ring}$ 
on fluxes, $V^2$, and differential phases predicted by disk models.
A smaller value of $R_{\rm in}$ means that while there is emission
from smaller radii, there will also be less emission
from gas at larger radii (for a given line-to-continuum ratio).  Thus
the line profile becomes broader as more flux is distributed
to the higher-velocity material at small radii.  The system becomes
less spatially resolved
at these high velocities (i.e., in the line wings), since the $L/C$ of the compact emission
has increased.  The emission becomes more resolved for lower velocities as the $L/C$
of this emission decreases (and the mean size thus shifts out toward the continuum 
ring).  Smaller values of $R_{\rm in}$ drive centroid offsets closer to zero,
and hence lead to smaller $\Delta \phi$.

Values of $R_{\rm in}$ larger than those shown in Figure
\ref{fig:disk_geometry} lead to even
narrower profiles of flux and $V^2$ versus wavelength, and larger
differential phase signatures.  For very large $R_{\rm in}$, the line
may become completely spectrally unresolved, in which case the $\Delta
\phi$ would become zero.  Since observed spectra for all targets show
spectrally resolve Br$\gamma$ line profiles (Figure
\ref{fig:spectra}), we do not consider values of $R_{\rm in}$ larger
than 0.05 AU in this modeling.

Because $R_{\rm ring}$ is also taken to be the outer radius of the gaseous disk,
it does have an impact on the observables beyond the normalization of
the continuum.  Larger values of  $R_{\rm ring}$ lead to larger centroid offsets between the
red and blue sides of the gaseous disk.  Moreover, larger values of
$R_{\rm ring}$ lead to a lower correlated continuum flux ($V_{\rm
  cont} F_{\rm cont}$),  meaning that the contribution of the symmetric
continuum component to the total measured differential phase signal is
decreased (see Equation \ref{eq:dpdisk}).  Thus, a larger radius of the continuum emission helps to
amplify any differential phase signal arising from the gaseous emission.
Note that $R_{\rm ring}$ is
determined directly from a fit of a ring model to continuum data, and
so is not actually a free parameter in the models.

\subsubsection{Disk Brightness Profile}
The disk brightness profile depends on the line-to-continuum ratio
($L/C$), which determines the normalization, and on $\alpha$, which
describes the radial profile.  Both have prominent effects on the
synthetic data (Figure \ref{fig:disk_brightness}).

Steeper brightness profiles, corresponding to higher values of $\alpha$, result in a higher
proportion of flux in the extreme velocities.  Hence large values
of $\alpha$ result in double-peaked profiles, while $\alpha$ values closer
to zero result in single-peaked profiles.  Models with shallower flux profiles will be more
spatially resolved, since more flux is found at larger radii.  Similarly, differential phases
deviate from zero more strongly for shallower flux profiles.  Note that the effects
of increasing $\alpha$ are similar to the effects of decreasing
$R_{\rm in}$.

Higher line-to-continuum ratios produce brighter lines in model
spectra (by definition).
Because of the higher proportion of flux in the compact gaseous component
relative to the continuum ring, models with higher line-to-continuum
ratios also produce less resolved (higher) $V^2$ within the emission line.
Higher line-to-continuum ratios mean that the centroid offsets between different
velocity components of the gaseous disk have less dilution from the continuum
ring (which has no centroid offset versus wavelength).  Thus we find larger differential
phase signatures for higher line-to-continuum ratios.

\subsubsection{Disk Viewing Geometry}
The position angle has no effect on the total flux emitted, although it may impact the
modeled $V^2$ and $\Delta \phi$, since models (and data) are only ``observed'' 
over a limited range of baseline position angles.  If the baseline PA is aligned with
the disk PA, then the system will appear more resolved, and differential phase signals
will be larger.  If the baseline PA is orthogonal to the disk PA,  the system will
be less resolved, and we will observe no differential phase deviations from zero.

Face-on disks show no radial velocity gradients, and near-to-face-on disks show
narrow emission features.  For sufficiently low inclinations, all of the disk emission
fits into a single spectral channel in our synthetic spectra. While the emission
in this channel will be more compact (higher $V^2$) than the surrounding 
continuum, such a model produces no differential phase signature since the
average centroid offset of the entire disk is zero.  In contrast, higher inclinations
mean broader emission lines, which lead to $V^2$ and $\Delta \phi$ signatures
in multiple channels.  The magnitude of these signatures depends on whether
the disk PA is aligned with the baseline PA,  and on the inclination.
In Figure \ref{fig:disk_viewing}, the gaseous disk becomes
less resolved with increasing inclination, since the system PA is not 
perfectly aligned with the baseline PA.

With a single baseline, we do not claim that we can constrain the
position angle or inclination of the disk.  However, we allow these
parameters to vary in the models to reflect potential misalignments
between the baseline $PA$ and the disk.

\subsection{Magnetospheric Infall/Outflow Models \label{sec:infall}}
We also consider models that include gaseous outflows or inflows
interior to the dust sublimation front.  We allow these outflows to
extend to scales larger than the dust sublimation radii.
As with the disk model considered in \S \ref{sec:diskmods}, we include a ring
of continuum emission. For simplicity here,
we assume the position angle and inclination of the ring are zero.  Note that
the ring geometry is not coupled to the morphology of the gas.

We assume an infall/outflow cone with an
opening angle of $5^{\circ}$.   The cone has a position angle, $PA$, and an
inclination with respect to the plane of the sky, $\phi$.  We allow the infall/outflow
to extend from an outer radius, $R_{\rm out}$ to an inner radius,
$R_{\rm in}$.  In contrast to the disk model considered above, $R_{\rm
  out} \ne R_{\rm ring}$ here.

The velocity of material in this cone is described as
a radial power-law:
\begin{equation}
v_{\rm obs}(R) = v_{\rm in} \left(\frac{R}{R_{\rm in}}\right)^{-\beta} \sin \phi.
\label{eq:outflow_vr}
\end{equation}
Here, the velocity of material at the inner edge of the infall/outflow
structure, $v_{\rm in}$, is chosen to produce a specified linewidth of
the emission,
\begin{equation}
\Delta v = v_{\rm in} \sin \phi.
\label{eq:outflow_vin}
\end{equation} 

Examination of Equations \ref{eq:outflow_vr} and \ref{eq:outflow_vin} shows that the
velocity profile depends on $\Delta v$, $R_{\rm in}$, and $\beta$, but not directly
on $\phi$.   The geometry of the outflow of the sky depends on $PA$, as well
as on $R_{\rm in}$ and $R_{\rm out}$, but again not explicitly on $\phi$.  We thus
fix $\phi=45^{\circ}$ in our models.

We include another cone, reflected through the
origin, with the same velocity profile multiplied by $-1$.  Thus, the
model includes a bipolar infall/outflow structure within the
dust sublimation radius.

The brightness distribution of the infall/outflow cones is
\begin{equation}
B_{\rm infall/outflow}(R) =  B_{\rm in} \left(\frac{R}{R_{\rm in}}\right)^{-\alpha},
\end{equation}
where $B_{\rm in}$ is chosen to reproduce a specified line-to-continuum ratio.
As above, $L/C$ is defined as the total, integrated flux of the gaseous emission
over the total flux of the continuum.
Note that $\alpha$ rolls all information about the temperature and surface
density profile of the infall/outflow structure into a single parameter.
Finally, we include as a free parameter a factor by which the flux in 
one of the two cones or ``poles'' or the infall/outflow may be scaled.
We denote this factor as $f_{\rm a}$, since it represents an asymmetry in the model.

\begin{figure}[tbp]
\plotone{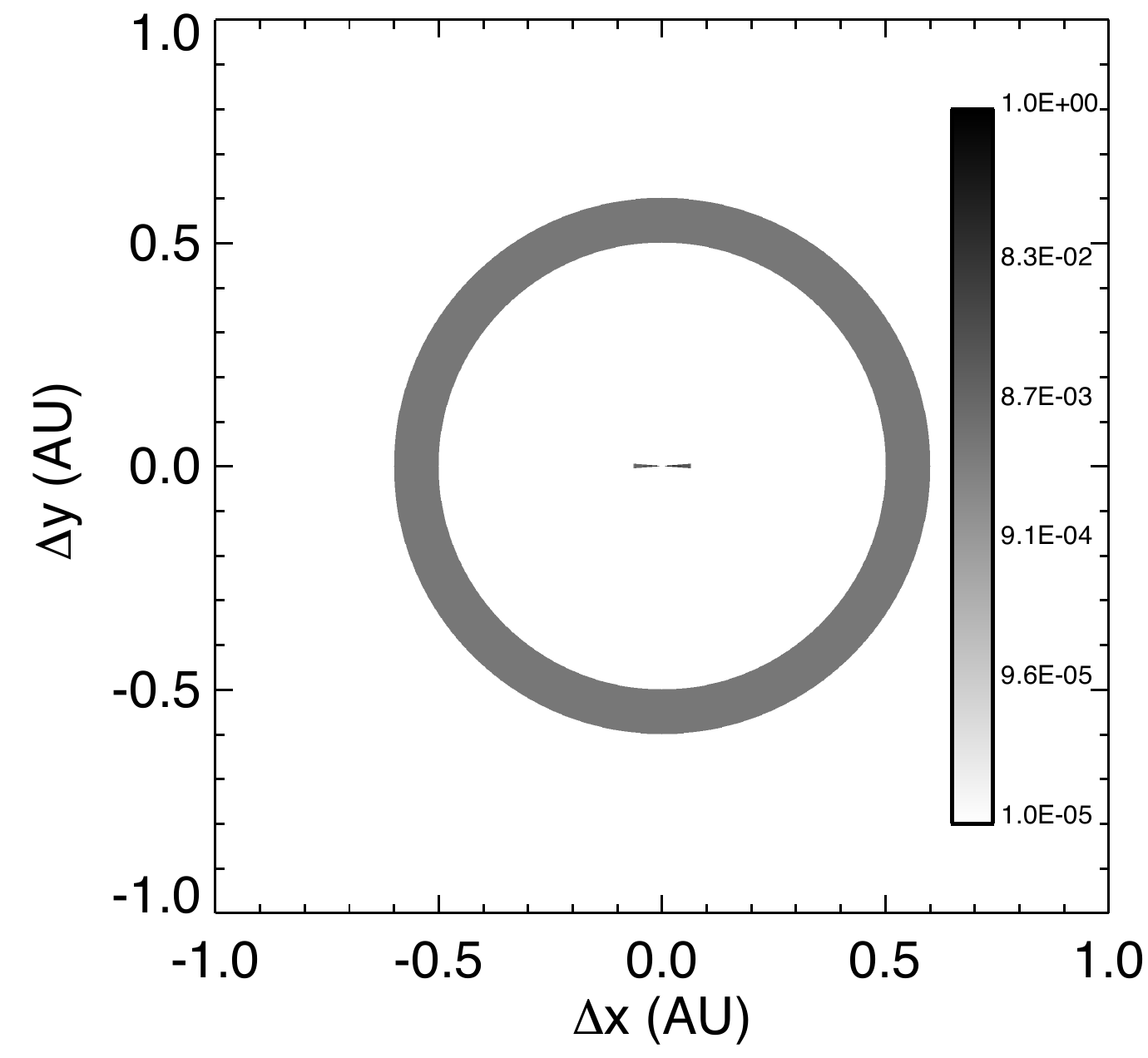}
\caption{Synthetic, velocity-integrated image, computed
for the infall/outflow model described in \S \ref{sec:infall} assuming
 $R_{\rm in}=0.01$ AU, $R_{\rm out}=0.1$ AU,
$PA=0$, $\alpha=2$, $\beta=2$,
$\Delta v=250$ km s$^{-1}$, $L/C=0.5$, and $f_{\rm a}=0.5$. 
The image has been normalized, and is shown with a logarithmic
stretch.  Note that the star is not included in this synthetic image.
\label{fig:infallimage}}
\end{figure}

\begin{figure}[tbhp]
\plotone{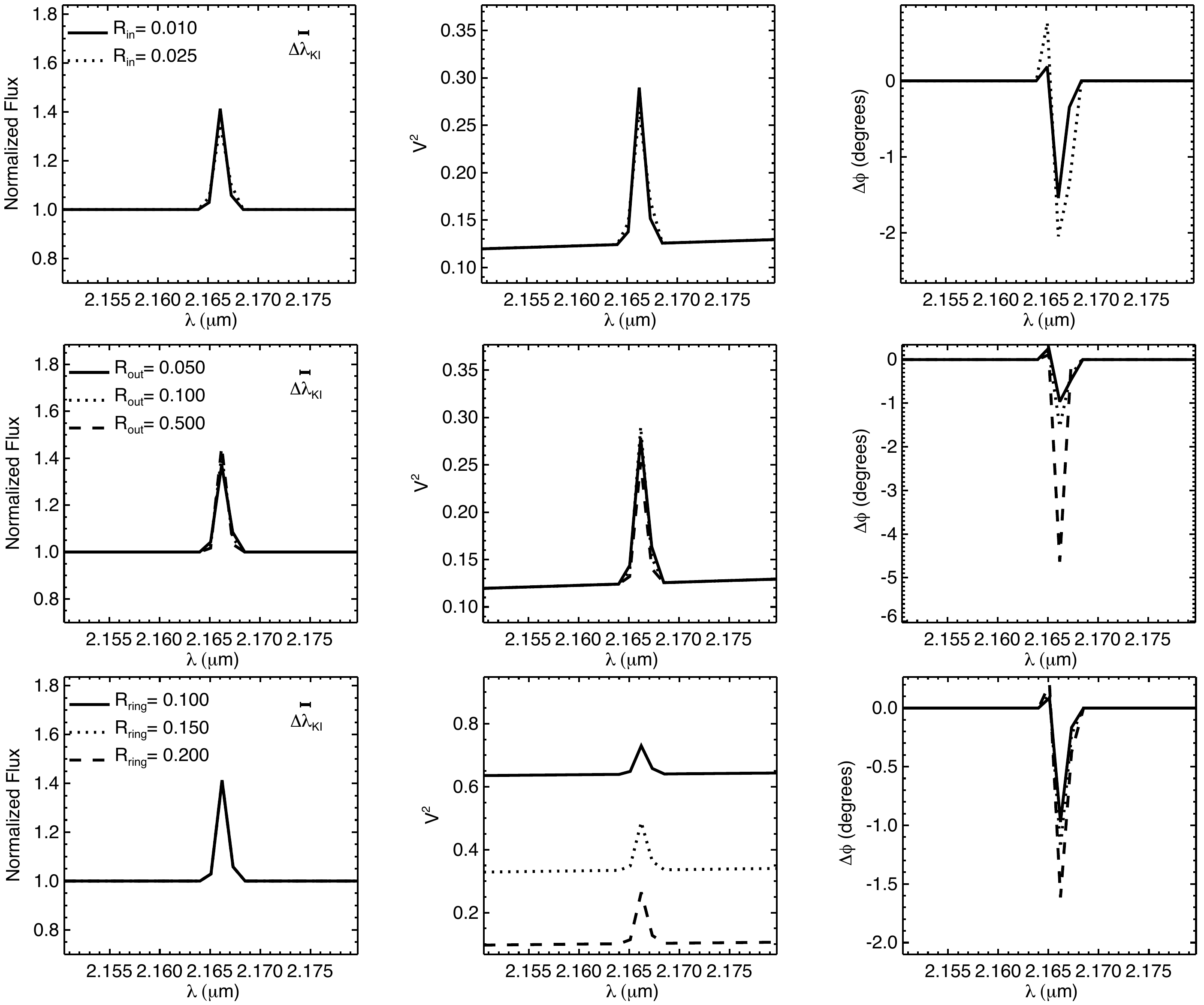}
\caption{Synthetic fluxes, $V^2$, and differential phases ($\Delta \phi$) computed
for the infall/outflow model described in \S \ref{sec:infall}.   The
fiducial model assumes $R_{\rm in}=0.01$ AU, $R_{\rm out}=0.1$ AU,
$PA=0$, $\alpha=2$, $\beta=2$,
$\Delta v=250$ km s$^{-1}$, $L/C=0.5$, $f_{\rm a}=0.5$, and $d=140$ pc. 
Here we vary $R_{\rm in}$, $R_{\rm out}$, and $R_{\rm ring}$ to
demonstrate the effects of these parameters on the synthetic data.
Note, however, that $R_{\rm ring}$ is not actually a free parameter in
our infall/outflow models.
\label{fig:infall_geom}}
\end{figure}

\begin{figure}[tbp]
\plotone{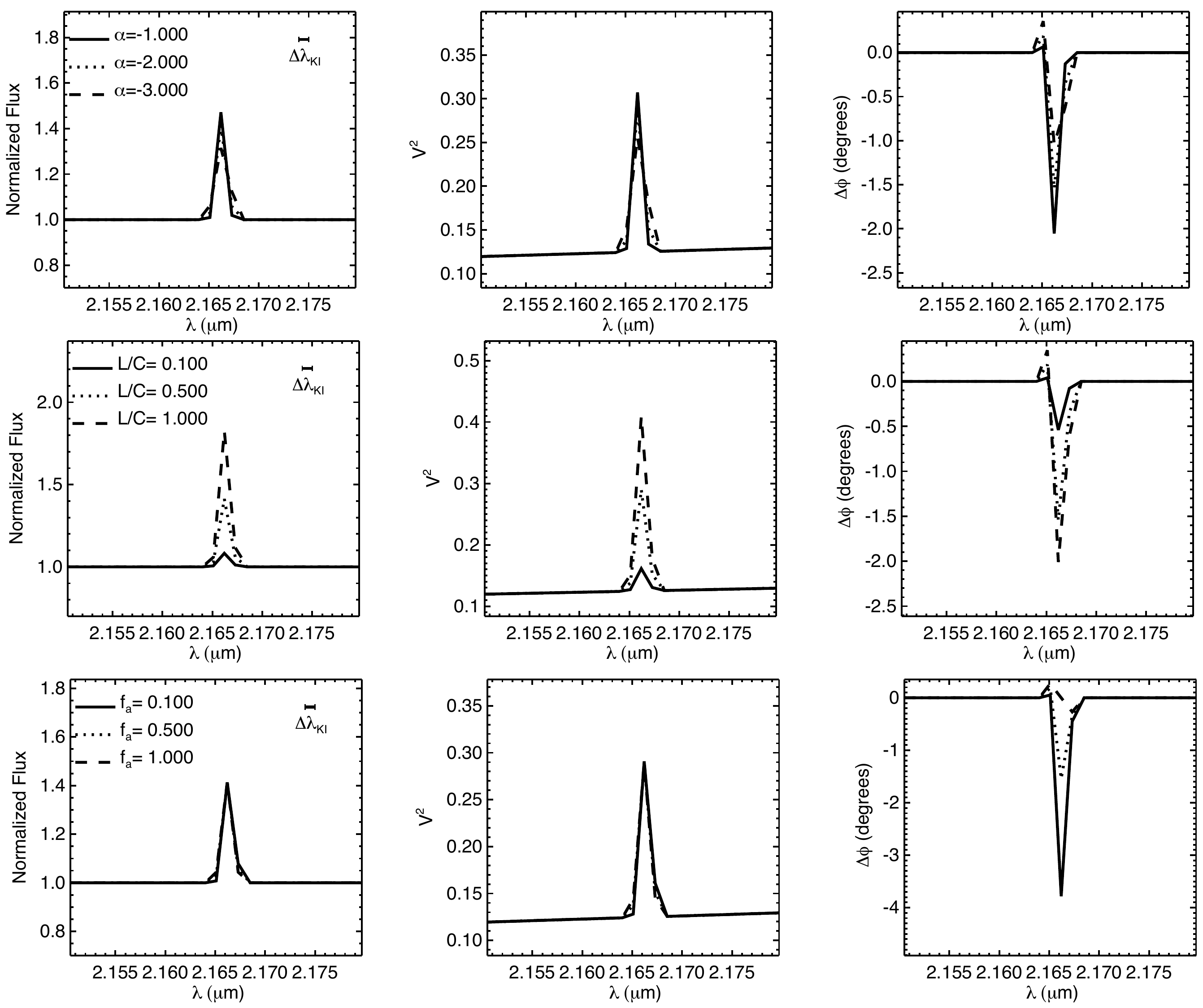}
\caption{Synthetic fluxes, $V^2$, and differential phases ($\Delta \phi$) computed
for the infall/outflow model described in \S \ref{sec:infall}.   The
fiducial model is as described in Figure \ref{fig:infall_geom}, and we
vary $\alpha$, $L/C$, and $f_{\rm a}$  here.
\label{fig:outflow_brightness}}
\end{figure}

\begin{figure}[tbp]
\plotone{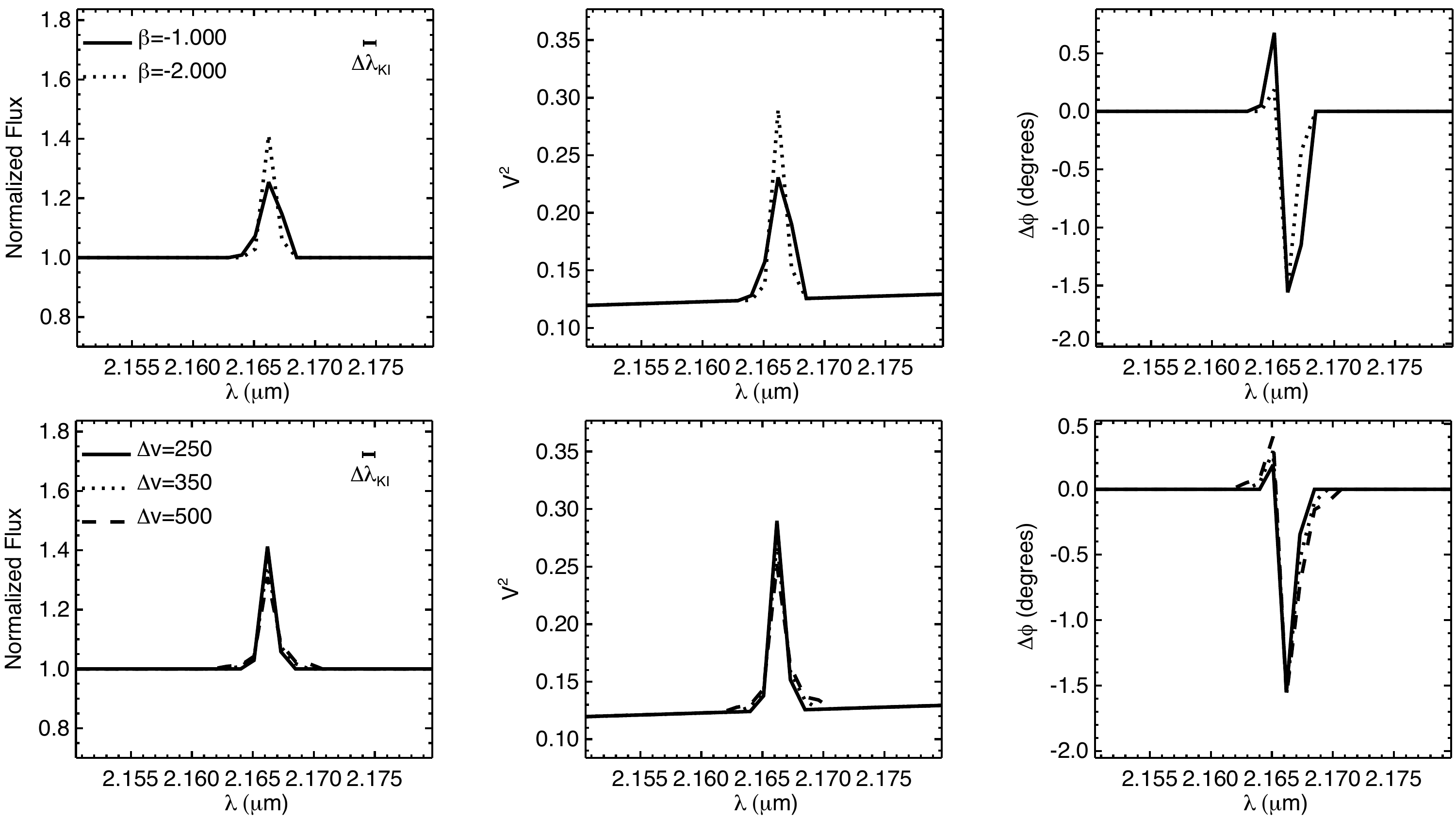}
\caption{Synthetic fluxes, $V^2$, and differential phases ($\Delta \phi$) computed
for the infall/outflow model described in \S \ref{sec:infall}.   The
fiducial model is as described in Figure \ref{fig:infall_geom}, and we
vary $\beta$ and $\Delta v$ here.
\label{fig:outflow_velocity}}
\end{figure}

\begin{figure}[tbp]
\plotone{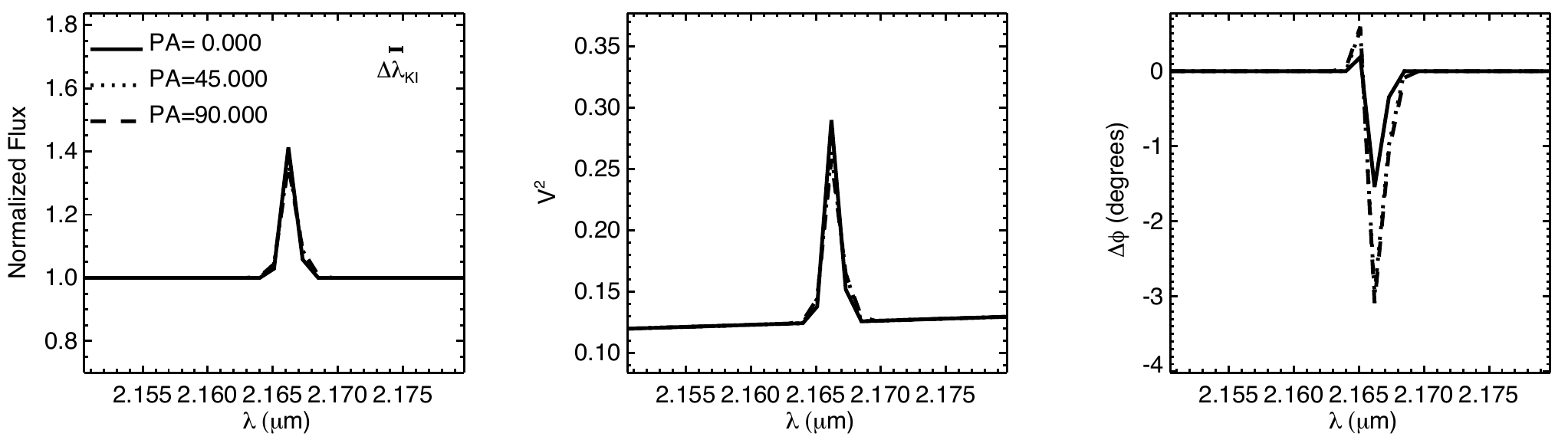}
\caption{Synthetic fluxes, $V^2$, and differential phases ($\Delta \phi$) computed
for the infall/outflow model described in \S \ref{sec:infall}.   The
fiducial model is as described in Figure \ref{fig:infall_geom}, and we
vary $PA$ here.
\label{fig:outflow_viewing}}
\end{figure}

An image of an example infall/outflow model is shown in Figure \ref{fig:infallimage}.
Different models may be generated by varying the free parameters
$R_{\rm in}$, $R_{\rm out}$, $PA$, $\alpha$, $\beta$, $\Delta v$, $L/C$, and $f_{\rm
  a}$.  We now summarize the effects of each of these parameters on
synthetic fluxes, $V^2$, and differential phases, and illustrate these effects
in Figures \ref{fig:infall_geom}--\ref{fig:outflow_viewing}.

\subsubsection{Infall/Outflow Geometry}
The inner and outer radii of the infall/outflow structure have small
effects on modeled fluxes and $V^2$, but can have a larger effect on
the differential phases.  The larger effects on $\Delta \phi$ stem
from the greater sensitivity of differential phases to compact
structure.  Since centroiding accuracy is better than the angular
resolution (by a factor of the S/N), smaller source structure can be
constrained.

A larger inner radius means that the highest-velocity material is
located farther from the star, and so covers more area. The relative
amount of flux in the high-velocity component of 
the outflow is thus increased, which leads to a flatter line profile.
Since the line-to-continuum ratio is increased for high velocity
material, but decreased for low-velocity gas, the $V^2$ versus
wavelength also flattens.  Larger values of $R_{\rm in}$ drive larger 
centroid offsets, and hence lead to larger $\Delta \phi$.
Similarly, larger values of $R_{\rm out}$ lead to (somewhat) more
extended emission, and hence to larger centroid offsets.

As for the disk model considered above, a larger value of $R_{\rm
  ring}$ leads to less correlated flux from the continuum component.
This decreases the contribution of the continuum to the
observed differential phases, and will amplify any differential phase
signatures coming from the infall/outflow component.

\subsubsection{Infall/Outflow Brightness Profile}
The effect of $\alpha$ on the observables is similar to the effects
described above in the context of disk models.  The line profile
becomes less peaked, the $V^2$ versus wavelength also become less
peaked, and the differential phase signatures get smaller as the
brightness profile of the model is steepened.
The dependence of the synthetic data on $L/C$ for the infall/outflow
model is the same as for the disk model described above.

Decreasing $f_{\rm a}$ increases the flux ratio between the blue and
red sides of the infall/outflow structure.  This manifests itself
through slight asymmetries in the synthetic flux and $V^2$ versus
wavelength. The effect on synthetic differential phases is more
pronounced, since these are more sensitive to the differences in 
emission on such compact scales.

\subsubsection{Infall/Outflow Velocity Structure}
The effect of $\beta$ on the observables similar to, but not degenerate with, the
effects of $\alpha$.  A steeper velocity profile means that more of
the low-velocity gas is found at smaller stellocentric radii, where
the emission is stronger.  Thus, a higher value of $\beta$ is similar to a smaller value of
$\alpha$.  However, comparison of this Figures \ref{fig:outflow_brightness} and
\ref{fig:outflow_velocity} demonstrates that the effects on the synthetic
data are sufficiently different so as to be distinguishable.

As $\Delta v$ is increased the line profile becomes broader and flatter (flatter because 
the line-to-continuum ratio is
held fixed for a given parameter study). The $V^2$ versus wavelength also becomes broader and flatter.  Examination of
Equations \ref{eq:outflow_vr} and \ref{eq:outflow_vin} shows that the physical effect of
increasing $\Delta v$ is to increase the radial velocity at the inner edge of the infall/outflow structure.
This, in turn, causes the radius where the infall/outflow reaches systemic velocity to move outward.
A larger offset of the near-zero velocity material leads to a larger differential phase signature.

\subsubsection{Infall/Outflow Viewing Geometry}
The effect of $PA$ on the observables is similar to the effects
described above in the context of disk models.  
Since the differential phases are sensitive to structure on smaller
scales than the $V^2$, and the infall/outflow model generates most of
its flux on compact scales, it is not surprising the $\Delta \phi$ is
more sensitive than $V^2$ to $PA$ changes.  For the definition of our
model, $\phi$ has no direct effect on the synthetic data.

\subsection{Model Fitting}
We compute grids of models using the parameter values shown in Figures
\ref{fig:disk_geometry}--\ref{fig:outflow_viewing}.
Grids with more finely-sampled parameter values are not possible 
given the computational time needed to compute these models.
Our disk model grid requires 1--2 weeks,
while our infall/outflow model grid (which has more free parameters) requires
over 3 weeks to run on a fast desktop computer.

After computing synthetic fluxes, $V^2$, and $\Delta \phi$ values for
grids of both disk and infall/outflow models, we compute 
the reduced $\chi^2$ residuals between these and the observed quantities.
The total $\chi^2$ is given by
\begin{equation}
\chi^2_{\rm tot} = \sqrt{(\chi^2_{\rm flux})^2 + (\chi^2_{\rm V^2})^2 +
 (\chi^2_{\rm \Delta \phi})^2}.
\end{equation}
Finally, we minimize $\chi^2_{\rm tot}$ to determine the ``best-fit'' model.
With such a sparse grid of models, we can not claim that our
``best-fit'' model is the true, absolute minimum of the $\chi^2$
surface, as opposed to a deep local minimum. Furthermore, we can not
give rigorous error intervals on the fitted parameters, since we have
not adequately sampled the $\chi^2$ surface.

\begin{figure}[tbhp]
\plotone{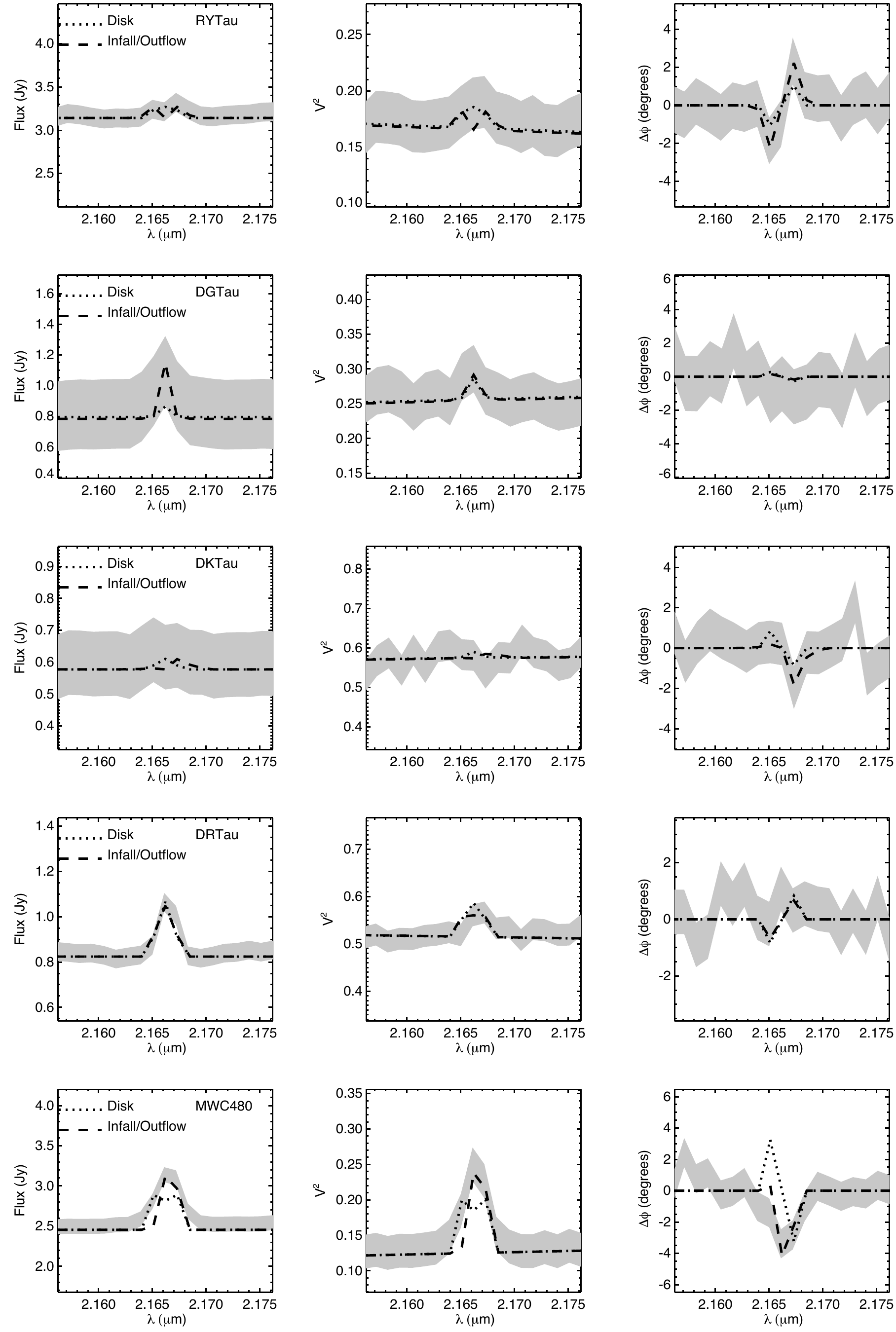}
\caption{Observed fluxes, $V^2$, and $\Delta \phi$ values (gray
  regions indicate 1-$\sigma$ confidence intervals), and synthetic data
for best-fit disk (dotted curves) and infall/outflow (dashed curves) models.
\label{fig:modfits}}
\end{figure}

The best-fit models are illustrated in Figure \ref{fig:modfits}.
Reduced $\chi^2$ values, and parameters of the best-fitting models, are listed in Table
\ref{tab:results}.

\section{Results and Analysis \label{sec:res}}

\subsection{Accretion properties inferred from Br$\gamma$ Spectra \label{sec:brgspec}}
Before discussing the distribution of circumstellar matter for our sample,
we begin by analyzing our Br$\gamma$ spectra and estimating accretion luminosities
using tools that have been developed and used previously.  These accretion
luminosities will provide some basis for comparison against the circumstellar
properties.

In Table \ref{tab:brg} we list the equivalent
widths (EWs) of the Br$\gamma$ lines measured for our sample, after removal of 
the stellar components of the spectra using the procedure described in 
\S \ref{sec:ratios}.  Following \citet{EISNER+07c}, we use these equivalent
widths in conjunction with literature photometry and
extinction estimates to determine the Br$\gamma$
line luminosities.   

These line luminosities are then converted into
accretion luminosities using an empirically-determined relationship
from \citet{MHC98,MCH01}.   This relationship is determined by
comparing Br$\gamma$ line luminosities to accretion luminosities
determined by fitting shock models to UV data.
Note, however, that this relationship was determined for
mostly solar-type T Tauri stars.  While the relationship has been
extended to $\sim 3$ M$_{\odot}$ stars \citep{CALVET+04}, it has not been
calibrated at high stellar masses and may break down for targets like
MWC 1080.  In fact, we argue below and elsewhere \citep[e.g.,][]{EISNER+04}
that this object may not undergo magnetospheric accretion, in
which case the relationship from \citet{MHC98,MCH01} is almost
certainly invalid.

The accretion luminosity can be converted into an accretion rate:
$\dot{M} \approx L_{\rm acc} R_{\ast} / GM_{\ast}$.  Since Br$\gamma$
is in absorption for V1057 Cyg, neither $L_{\rm acc}$ nor $\dot{M}$
is meaningful in this case.  For AS 353 and V1331 Cyg, we do not
have reliable estimates of stellar parameters, and so we can not
estimate $\dot{M}$.  

Table \ref{tab:brg} shows that our sample spans a wide range in Br$\gamma$
luminosity, and hence in accretion luminosity and mass accretion rate.
Measured EWs (and accretion luminosities and mass accretion rates)
are generally within a factor of two of previous measurements
\citep{NCT96,FE01,CALVET+04,EISNER+07c}.  Given the large variations in Br$\gamma$
EW over multiple epochs observed by previous investigators 
\citep[e.g.,][]{NCT96,EISNER+07c}, our measurements seem compatible
with the previous results.

\subsection{Distribution of Br$\gamma$ Emission \label{sec:brgres}}

For many objects in our sample, both the disk and infall/outflow
models considered above provide reasonable fits to the data
(Table \ref{tab:results}), and so we can not distinguish between
the two models, at least based on the available data.  However,
there are several sources for which we can distinguish. For 
MWC 480, MWC 863, and MWC 275, the $\chi_r^2$
values for fits of the disk model are substantially higher than
those of infall/outflow model fits.   These objects all show
strong Br$\gamma$ emission, and have been observed
at high S/N. It seems likely that higher S/N observations
of some fainter targets in our sample--especially those exhibiting
bright Br$\gamma$ emission--would also show them to
be better described by infall/outflow models.

To understand why the data for these objects are fitted better by
infall/outflow models, consider the following argument.
If the Br$\gamma$ line is spectrally resolved--which is easiest to
determine for bright objects with strong emission--then one
expects to see differential phase signatures as long as the
gaseous emission is not very compact.  If the gas is very compact,
this leads to double-peaked line profiles for disk models,
contradicting the observations.  To ensure single-peaked 
line profiles, the brightness profiles of disk models must be
shallow.  However, this produces differential phase signals
larger than observed.  
Infall/outflow models, on the other hand, can produce single-peaked line
profiles without large centroid offsets.  For example, if one posits an infall pointed nearly
toward the observer, a lot of flux could come from near-systemic velocities without
creating a large centroid offset (on the sky) of different velocity components.

For fainter sources, it is difficult
to apply this argument, because of the lower signal-to-noise
in both flux and differential phase measurements.  Noisier fluxes
make it harder to rule out models predicting spectrally unresolved
emission (which also predict no $\Delta \phi$ signature), while
noisy $\Delta \phi$ measurements make it harder to rule out
large centroid offsets with wavelength of the model emission.
However, it is likely that observations of (some of) these fainter sources
with higher S/N would find a preference for infall/outflow
models .

%The data for one target are fitted marginally better with 
%disk models than with infall/outflow models (Table \ref{tab:results}):
%MWC 758. This target a fairly bright, but shows relatively weak Br$\gamma$ emission.
%As such, we do not put great weight in the difference in $\chi_{\rm r}^2$ for disk
%and outflow models for this object.

We note that the orientation of the KI baseline with respect to the
disk position angle can be estimated in a few cases.  For MWC 275, MWC
480, and MWC 758, estimated disk PAs \citep{TANNIRKULAM+08,EISNER+04}
are nearly orthogonal to the baseline PA.  This supports our
interpretation of the data for MWC 480 and MWC 275 in terms of outflow
models.  In contrast, for MWC 1080, the estimated disk PA
\citep{EISNER+04} is nearly aligned with the KI baseline, suggesting
that the Br$\gamma$ observations may be sensitive to rotating gas in this source.

\section{Discussion}

\subsection{Br$\gamma$ as a Tracer of Magnetospheric Accretion \label{sec:accmods}}
In \S \ref{sec:brgres}, we argued that the spectra, $V^2$, and differential phases
measured for our sample are more compatible with infall/outflow models than
with disk models, at least in some cases.
Somewhat less direct evidence also argues against disk models for some of our targets.
Low-mass young stars are thought to accrete
material not via a boundary layer between star and accretion disk, but rather
through funnel flows originating near the magnetospheric radius and then
following stellar magnetic field lines to high-latitude regions of the 
star \citep[e.g.,][]{KONIGL91,HARTMANN98}.  This magnetospheric accretion picture provides
a natural mechanism for truncating circumstellar disks and hence allowing direct
illumination of the inner edge.  This, in turn, would lead to the puffed-up inner
disk walls evinced by continuum measurements of the inner regions
of disks around T Tauri and Herbig Ae stars 
\citep[][and references therein]{MILLAN-GABET+07}.

While our data are often fitted better with infall/outflow models, we can not
distinguish between infall and outflow models directly with the spatial
and kinematic constraints provided by our observations.  However, we
can compare our results with size scales of gaseous emission
predicted by various physical infall or outflow models.  Matter falling in along stellar magnetic 
field lines will glow brightly near to the stellar surface, where it converts most of its 
gravitational potential energy to radiation, and so we would expect to see Br$\gamma$ 
emission from very close to the stellar surface in this case \citep[e.g.,][]{MHC98}.  
For models where the Br$\gamma$ emission arises in hot, magnetically-launched 
winds, we might see emission close to the magnetospheric radius at hundredths to 
tenths of an AU \citep[as in the X-wind model; e.g.,][]{SHU+94} 
or in the range of an AU or more \citep[as in the disk-wind model; e.g.,][]{KP00}.

Our modeling suggests that the Br$\gamma$ emission in most objects
arises in material that extends very close to the central star ($\la
0.01$ AU; Table \ref{tab:results}), consistent with expectations of
magnetospheric infall models.  This conclusion is supported for most
sources by the very small average sizes of the Br$\gamma$ emission
found from simple geometric fits to the data (Table
\ref{tab:brgsizes}).

However, for some targets, we infer somewhat larger inner radii and/or
shallow brightness profiles for the Br$\gamma$ emission. Somewhat
larger sizes are also inferred for the average Br$\gamma$ emission
from these targets ($\sim 0.05$--0.2 AU; Table
\ref{tab:brgsizes}).  Evidently, much of the Br$\gamma$ emission from
these objects is found on scales larger than expected from accretion columns.
X-wind or disk wind models predict more Br$\gamma$ emission from
larger radii, potentially consistent with the inferred size scales and brightness
profiles for these sources.  

Based on these arguments it seems likely that in the sample of objects
with steeper brightness profiles, Br$\gamma$ emission traces
magnetospheric accretion.  For sources with shallower brightness profiles,
Br$\gamma$ emission traces:
infalling material that has a shallow radial brightness profile; outflowing material
that is magnetospherically launched from small radii;  or a combination of 
both infalling and outflowing material.  In the latter case, the infalling material could
produce very compact emission while more extended winds could ``fill in'' the
emission profile as larger radii.  

Further support for magnetospheric accretion as the origin for (some of) the
Br$\gamma$ emission comes from
analysis of line profiles observed at higher spectral (but lower
spatial) resolution.  While most of our targets exhibit H$\alpha$
emission showing P Cygni profiles of winds \citep[e.g.,][and
references therein]{AVD05,NCT96}, the Br$\gamma$ line profiles are
more symmetric and/or blueshifted, and often show a ``blue shoulder''
\citep{NCT96,FE01,EISNER+07c}.  These properties are all consistent
with infalling material rather than winds \citep{NCT96}.

\subsection{Trends with Stellar Properties \label{sec:stellar}}
While magnetospheric accretion is favored for low-mass stars, previous
investigators have argued that disk accretion may be a better description in
Herbig Be stars \citep[e.g.,][]{EISNER+04}.  Thus, for MWC 1080, the one Herbig
Be star in our sample, one might expect the disk model considered above
to fit the data well.  Note that the gaseous disk described in \S \ref{sec:diskmods}
could correspond physically to the inner accretion disk expected for a boundary layer 
accretion scenario where disk material in Keplerian rotation shocks when it hits the
slower-rotating stellar equator \citep[e.g.,][]{LP74}.   

Disk models fit the data for this source approximately as well as outflow
models.  Furthermore, the orientation of the KI baseline along the
disk major axis lends credence to the hypothesis that the Br$\gamma$
we observe traces disk rotation.  Thus, our current study supports the idea
that Herbig Be stars may accrete material through
geometrically thin disks rather than magnetospheric accretion columns.

\subsection{Trends with Br$\gamma$ Line Luminosities \label{sec:brgcorr}}
Our sample clearly exhibits a range in Br$\gamma$ line-strength (Figure \ref{fig:spectra}),
which correlates with Br$\gamma$ line luminosity and accretion luminosity
(Table \ref{tab:brg}).  We are thus in a 
position to explore whether the circumstellar properties of the Br$\gamma$
emission depend on the overall strength of the emission.

Beyond stronger line-to-continuum ratios, it is unclear what we would expect
the effects of stronger Br$\gamma$ emission (and by extension accretion luminosity
or rate) to be on the observed data and models.  If Br$\gamma$ emission strength
was correlated with the emission morphology--e.g., if stronger Br$\gamma$ emission
tended to trace infall while weaker emission traced outflow from larger stellocentric radii--then
we might expect trends in the data.  

\begin{figure}[tbp]
\plotone{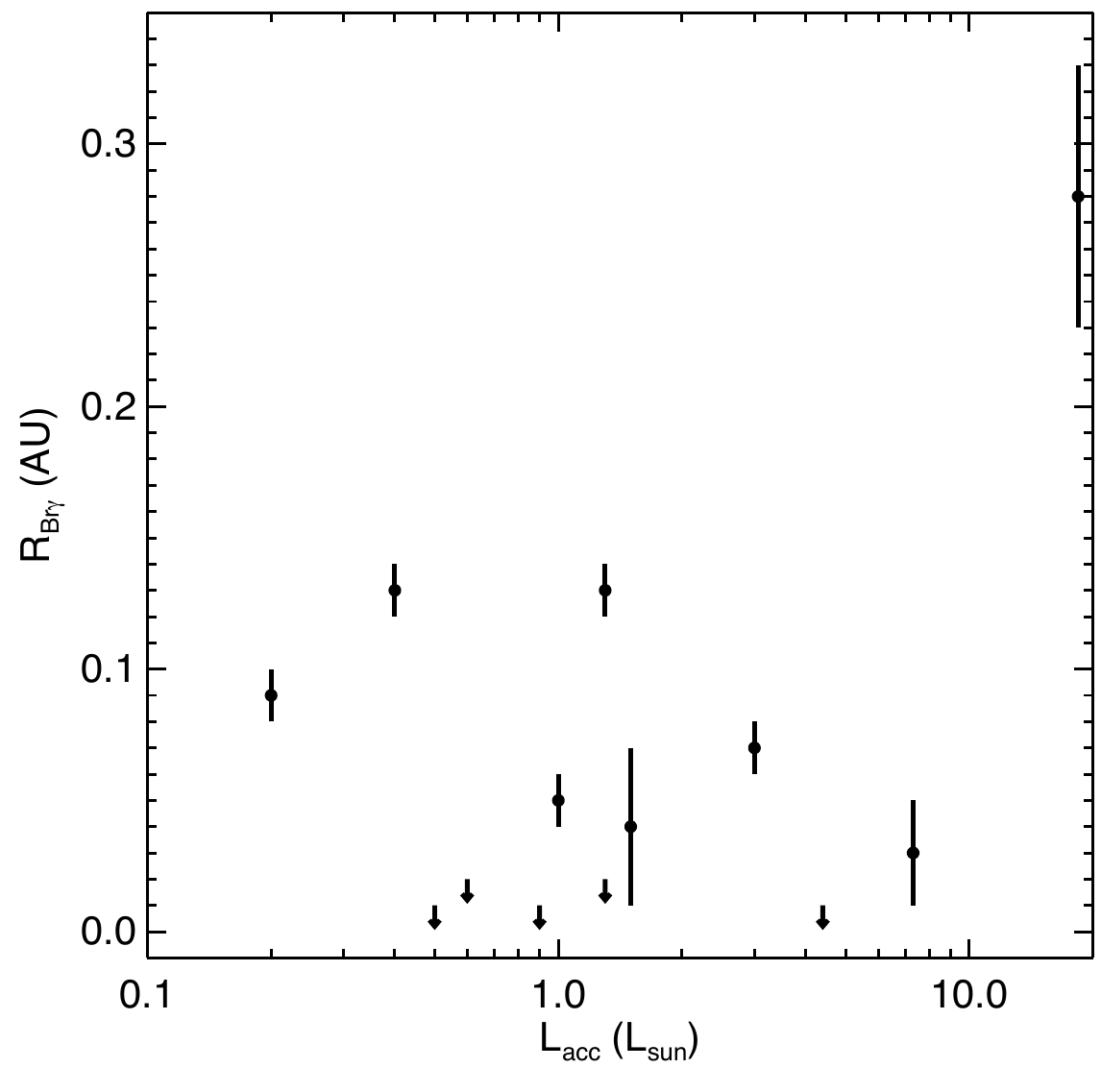}
\caption{Inferred average stellocentric radius of Br$\gamma$ emission (from Table
\ref{tab:brgsizes}) plotted against the accretion luminosities derived for these
objects (from Table \ref{tab:brg}).
\label{fig:lacc_brgsize}}
\end{figure}

The simplest way to search for trends is to look at inferred size of
the Br$\gamma$ emission versus the accretion luminosity.  Figure
\ref{fig:lacc_brgsize} shows no obvious correlation between the two.
Similarly, a comparison of Tables \ref{tab:results} and \ref{tab:brg}
shows no clear correlation between $L_{\rm Br\gamma}$ and $\alpha$.
Thus, there is no strong indication that Br$\gamma$ emission strength
determines the overall morphology.

\begin{figure}[tbp]
\plotone{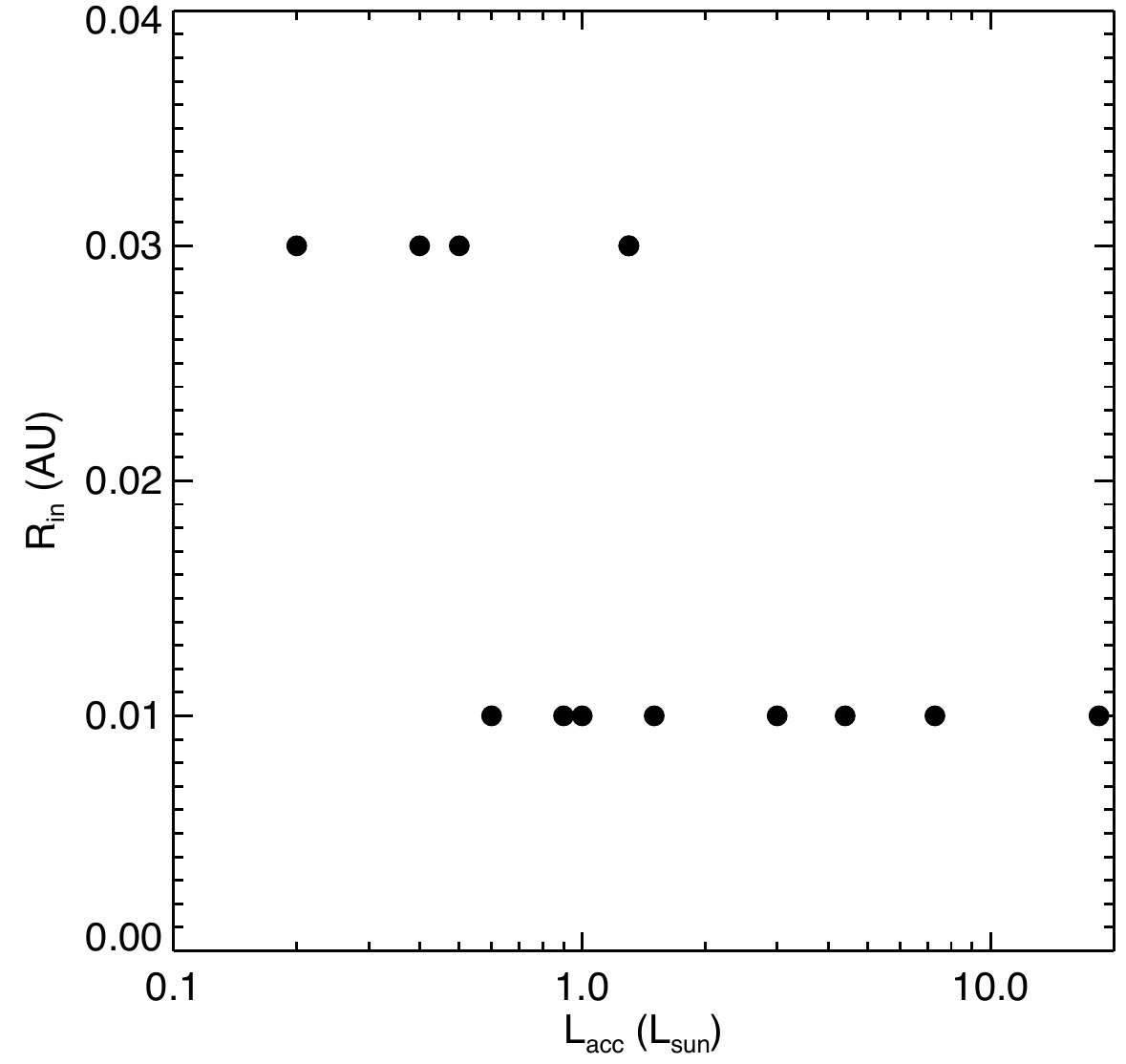}
\caption{Best-fit values of $R_{\rm in}$ for the infall/outflow models (from Table
\ref{tab:results}) plotted against the accretion luminosities derived for these
objects (from Table \ref{tab:brg}).  We only considered these two
values of $R_{\rm in}$ in our modeling, which is why this plot is quantized.
\label{fig:lacc_rin}}
\end{figure}

We do, however, see some (inverse) correlation between the inferred inner radius
of the Br$\gamma$ emission and the accretion luminosity.  Considering
best-fit parameters for infall/outflow model fits, and excluding
MWC 1080, which appears to be fitted better with a disk model, we plot
the inner radius of the infall/outflow inferred from our modeling
against $L_{\rm acc}$  in Figure \ref{fig:lacc_rin}.
This may indicate that sources that convert more gravitational energy
into Br$\gamma$ luminosity do so at preferentially smaller stellocentric radii.
For example, sources where Br$\gamma$ traces accreting flows will
exhibit the most compact emission, and may also produce higher
Br$\gamma$ luminosities.  Alternatively, the trend seen in Figure
\ref{fig:lacc_rin} could fit in with an X-wind scenario where Br$\gamma$
emission traces the innermost regions of an outflow: higher accretion 
rates--which mean more gravitational energy release--push the ``X-point''
to smaller stellocentric radii as accretion pressure pushes in against stellar
magnetic pressure. 

%Using stellar masses and radii, and accretion rates for our sample, and assuming
%a magnetic field strength of 2kG \citep[e.g.,][]{JVG03}, we calculate
%magnetospheric radii $\ga 0.03$ AU  for our sample.  
%However, observations of Herbig Ae stars have found magnetic field
%strengths approximately an order of magnitude smaller \citep[e.g.,][]{HUBRIG+07},
%and recent observations suggest that the dipole component of the magnetic
%field (which is relevant for magnetospheric accretion) may be a small fraction
%of the $\sim 2$ kG measured for the total field of many stars \citep{JOHNSKRULL09}.
%Thus we do not rule out the possibility that magnetospheric radii could
%lie at stellocentric radii less than 0.01 AU for some of the stars in our sample.

%However, it is unclear whether this correlation is physically meaningful,
%since it is not present if $L_{\rm acc}/L_{\ast}$ is plotted instead of $L_{\rm acc}$.

\subsection{Comparison with previous results \label{sec:comparison}}
As discussed in the introduction, \citet{KRAUS+08} observed five Herbig Ae/Be
sources with the VLTI interferometer, and obtained spatially and spectrally
resolved data indicating that the Br$\gamma$ emission arose on scales larger
than expected for magnetospheric accretion models.  Their derived sizes of
the gaseous emission were a few tenths of an AU for the Herbig Ae sources and 
a few AU for the Herbig Be objects.  

Fitting ring models to the Br$\gamma$ emission, as done by
\citet{KRAUS+08}, we typically find substantially smaller sizes (Table
\ref{tab:brgsizes}).   For MWC 275, the size we infer for
the Br$\gamma$ emission is $\sim 10$ times smaller than
that found by \citet{KRAUS+08}.  We note that the components of the
MWC 275 visibilities attributed to Br$\gamma$ by \citet{KRAUS+08} are all
consistent with unity, and we would therefore argue that their data
are consistent with the compact size for this object determined here.
Given the greater angular resolution of KI relative to VLTI, we are
more sensitive to emission on compact scales, and so it is not overly
surprising that we are able to better constrain Br$\gamma$ emission at
small stellocentric radii.

The modeling presented here suggests emission over a range of radii.
While some objects have average Br$\gamma$ emission sizes as large as
$\sim 0.1$ AU (Table \ref{tab:brgsizes}), 
the combination of spectrally resolved flux and $V^2$
profiles and small differential phases for these targets indicates some
Br$\gamma$-emitting gas on $\sim 0.01$ scales (Table
\ref{tab:results}).  Thus, all of our targets where Br$\gamma$
emission is observed appear to have some gas on very compact scales.

While \citet{KRAUS+08} argued that the Br$\gamma$ emission appeared to
trace more extended disk winds, our results and modeling belie this
conclusion.  Rather, we suggest that the Br$\gamma$ emission traces
accreting material in many, if not most, sources.  In others, Br$\gamma$ emission
may trace a combination of accretion and outflow, or perhaps compact
($\la 0.05$ AU) wind-launching regions.

\section{Conclusions and Future Work \label{sec:conc}}
We presented observations capable of resolving hydrogen gas on scales
as small as 0.01 AU around young stars.  These were the first such
observations of solar-type T Tauri stars, and extended the mass range
and sample size of previous studies at similar spectral (but somewhat
lesser spatial) resolution.

We showed that Br$\gamma$ emission is typically more compactly
distributed than the continuum emission around young stars.
In several objects, the bulk of the Br$\gamma$ emission is found on scales $\la 0.01$
AU.  In others, the average size of the Br$\gamma$ emission is $\sim
0.1$ AU, although modeling of our combined dataset shows evidence for
some Br$\gamma$ emission extending in to within a few hundredths of an AU of the
central stars.  For sources with very small average sizes of the
Br$\gamma$ emission, an origin of the Br$\gamma$ emission in accretion
flows appears to be the best explanation.  For objects with somewhat
more extended Br$\gamma$ distributions, the emission probably traces
the innermost regions of magnetospherically
launched winds, or perhaps some combination of infall and outflow.

No obvious trends are seen in the average size of the Br$\gamma$
emission versus accretion luminosity.  However,
there appears to be some inverse correlation
between the inner radius for our best-fit models and the Br$\gamma$
emission strength.  Such a correlation may indicate a relationship
between the physical origin of the Br$\gamma$ emission and its
luminosity.

While all of our observations are compatible with infall/outflow
models, the data for the most massive star in our sample, MWC 1080,
appear equally consistent with a Keplerian disk model.  Previous
investigators have argued that this object may have a dense, gaseous,
inner disk that prevents direct stellar irradiation of dust near the
sublimation radius.  Our modeling of this source is consistent with a
disk origin of the Br$\gamma$ emission.

While we observed V1057 Cyg, we did not discuss it at length here.
V1057 Cyg is the one target in our sample
where Br$\gamma$ appears in absorption, 
rather than emission.  As an FU Ori star, the luminosity of this
source is dominated by accretion energy released in the disk midplane
\citep[e.g.,][]{HHC04}, and so 
most lines should appear in absorption \citep[e.g.,][]{CHK91}.
We plan to discuss this source 
in detail in a future paper targeting FU Ori sources.  

We also plan to extend the analysis presented above to other spectral
regions. Although we focused on Br$\gamma$ here, our experimental
setup included the CO overtone  bandheads as well as regions of
significant opacity from water vapor.  In fact, several objects
discussed here also show interesting spectral features in
the region of the CO bandheads. Because the currently observed
$V^2$ and $\Delta \phi$ signatures associated with these features
are marginal, we postpone discussion to future
work, when we hope to have higher S/N data.

\vspace{0.3 in}

Data presented herein were obtained at the W. M. Keck Observatory,
from telescope time allocated to the National Aeronautics and Space 
Administration through the agency's scientific partnership with the California 
Institute of Technology and the University of California. The Observatory was 
made possible by the generous financial support of the W. M. Keck Foundation. 
The authors wish to recognize and acknowledge the cultural role and reverence
that the summit of Mauna Kea has always had within the indigenous Hawaiian
community. We are most fortunate to have the opportunity to conduct 
observations from this mountain. The ASTRA program, which enabled the
observations presented here, was made possible
by funding from the NSF MRI grant AST-0619965.
This work has used software from 
NExSci at the California Institute of Technology.
The Keck Interferometer is funded by the National Aeronautics and
Space Administration as 
part of its Exoplanet Exploration program.

\bibliographystyle{apj}
\bibliography{jae_ref}

\clearpage
\begin{deluxetable}{lccccccc}
\tabletypesize{\scriptsize}
\tablewidth{0pt}
\tablecaption{Target and Calibrator Properties
\label{tab:sample}}
\tablehead{\colhead{Source} & \colhead{$\alpha$} 
& \colhead{$\delta$} & \colhead{$d$} & \colhead{Spectral Type} & 
\colhead{$m_{V}$} & \colhead{$m_{K}$} & \colhead{References} \\
 & (J2000) & (J2000) & (pc) & & & &}
\startdata
\multicolumn{8}{c}{Target Stars} \\
\hline
RY Tau & 04 21 57.409 & +28 26 35.56 & 140 & K1 & 10.2 & 5.4 & 1 \\
DG Tau & 04 27 04.700 & +26 06 16.20 & 140 & K3 & 12.4 & 7.0 & 2 \\
DK Tau A & 04 30 44.28 & +26 01 24.6 & 140 & K9 & 12.6 & 7.1 &  3 \\
DR Tau & 04 47 06.21 & +16 58 42.8 & 140 & K4 & 13.6 & 6.9 & 2 \\
MWC 480 & 04 58 46.266 & +29 50 37.00 & 140 & A2 & 7.7 & 5.5 & 4 \\ 
RW Aur A & 05 07 49.568 & +30 24 05.161 & 140 & K2 & 10.5 & 7.0 & 2 \\  
MWC 758 & 05 30 27.530 & +25 19 57.08 & 140 & A3 & 8.3 & 5.8 & 4 \\
AS 205 A & 16 11 31.402 & -18 38 24.54 & 160 & K5 & 12.1 & 6.0 & 6 \\
MWC 863 A & 16 40 17.922 & -23 53 45.18 & 150 & A2 & 8.9 & 5.5 & 5 \\
V2508 Oph & 16 48 45.62 & -14 16 35.9 & 160 & K6 & 13.5 & 7.0 & 6 \\
MWC 275 & 17 56 21.288 & -21 57 21.88 & 122 & A1 & 6.9 & 4.8 & 5 \\
AS 353 A & 19 20 30.992 & +11 01 54.550 & 150 & F8 & 12.5 & 8.4 & 7\\
V1057 Cyg & 20 58 53.73 & +44 15 28.54 & 600 & G5 & 11.7 & 6.2 & 8 \\
V1331 Cyg & 21 01 09.21 & +50 21 44.8 & 700 & G5 & 11.8 & 8.6 & 9 \\
MWC 1080 & 23 17 25.574 & +60 50 43.34 & 1000 & B0 & 11.6 & 4.7 & 4 \\
\hline
\multicolumn{7}{c}{Calibrator Stars} & Applied to: \\
\hline
HD23642 & 03 47 29.453 & +24 17 18.04 &  110 & A0V & 6.8 & 6.8 & DG Tau,DK Tau A,DR Tau,RW Aur A\\
HD23632 & 03 47 20.969 & +23 48 12.05 & 120 & A1V & 7.0 & 7.0 & DG Tau,DK Tau A,DR Tau,RW Aur A\\
HD 23753 & 03 48 20.816 & +23 25 16.499  & 104 & B8V & 5.4 & 5.7 & RY Tau,MWC 480,MWC 758 \\
HD 27777 & 04 24 29.155 & +34 07 50.73 & 187 & B8V & 5.7 & 6.0 & RY Tau,MWC 480,MWC 758 \\
HD31464 & 04 57 06.426 & +24 45 07.90 & 45 & G5V & 8.6 & 7.0 & DG Tau,DK Tau A,DR Tau,RW Aur A\\
HD139364 & 15 38 25.358 & -19 54 47.45 & 53 & F3V & 6.7 & 5.7 & AS 205 A \\
HD141465 & 15 49 52.297 & -17 54 07.007 & 43  & F3V & 6.8 & 5.9 & AS 205 A \\ 
HD144821 & 16 08 16.582 & -13 46 08.582 & 76 &  G2V  & 7.5 & 6.0 & AS 205 A,V2508 Oph \\ 
HD148968 & 16 32 08.085 & -12 25 53.910 & 146 & A0V & 7.0 & 7.0 & AS 205 A, V2508 Oph \\
HD 149013 & 16 32 38.133 & -15 59 15.12 & 41 & F8V & 7.0 & 5.7 & MWC 863 \\
HD 163955 & 17 59 47.553 & -23 48 58.08 & 134 & B9V & 4.7 & 4.9 & MWC 275 \\
HD 170657 & 18 31 18.960 & -18 54 31.72 & 13 & K1V & 6.8 & 4.7 & MWC 275 \\
HD183442 & 19 29 30.077 & +03 05 23.607  & & B7V & 8.1 & 8.4 & AS 353 A\\
HD192985 & 20 16 00.615 & +45 34 46.291 & 35 &  F5V& 5.9 & 4.8 & V1057 Cyg \\ 
HD195050 & 20 27 34.258 & +38 26 25.194 & 83 & A3V & 5.6 & 5.5  & V1057 Cyg \\
HD198182 & 20 46 53.060 & +47 06 41.502 & 185 & A1V & 7.8 & 7.8 & V1331 Cyg \\ 
HD219623 & 23 16 42.303 & +53 12 48.512  & 20 & F7V & 5.6 & 4.3 & MWC 1080 \\
\enddata
\tablerefs{(1) \citet{MUZEROLLE+03}; (2) \citet{WG01}; (3) \citet{MMD98}; (4) \citet{EISNER+04};
(5) \citet{MONNIER+06}; (6) \citet{EISNER+05}; (7) \citet{PGS03}; (8) \citet{HPD03}; (9) \citet{EISNER+07c}.  
Calibrator star distances are
based on Hipparcos parallax measurements \citep{PERRYMAN+97}.}
\end{deluxetable}

\begin{deluxetable}{lccc}
\tabletypesize{\scriptsize}
\tablewidth{0pt}
\tablecaption{Log of Observations
\label{tab:obs}}
\tablehead{\colhead{Source}
& \colhead{Date} & \colhead{$u$ (m)} & \colhead{$v$ (m)}}
\startdata
RY Tau & 2008 Novemver 18 & 56,56,56,56,46,46,45 & 57,57,58,58,71,71,71 \\
DG Tau & 2008 November 17 & 55,55,56,56,47,47,30,30 & 51,51,54,54,70,70,77,77 \\
DK Tau & 2008 November 17 & 56,56 & 56,56 \\
DR Tau & 2008 November 17 & 56,56,41,40 & 60,61,72,72 \\
& 2008 Novemver 18 & 56,56,41,40 & 60,61,72,72 \\
MWC 480 & 2008 Novemver 18 & 56,46,46 & 56,70,71 \\
RW Aur & 2008 November 17 & 56,56,56,56,55,33,32 & 55,55,58,58,59,77,78 \\
& 2008 Novemver 18 & 56,56,56,56,55,33,32 & 55,55,58,58,59,77,78 \\
MWC 758 & 2008 Novemver 18 & 56,56 & 58,59 \\
AS 205a & 2009 July 15 & 32,32,31,53,53,53,51,50 & 45,45,45,54,54,53,52,52 \\
MWC 863 & 2009 July 15 & 44,44 & 44,43 \\
V2508 Oph & 2008 April 25 & 35,34,31 & 50,50,49 \\
MWC 275 & 2009 July 15 & 52,52,50,50,54,54,50,49 & 51,51,49,49,53,52,49,48 \\
AS 353 & 2009 July 15 & 52 & 66 \\
V1057 Cyg & 2009 July 15 & 54,54,52,51,51,49,49 & 56,57,60,61,61,64,64 \\
V1331 Cyg & 2009 July 15 & 44,43,40,39 & 67,67,71,71 \\
MWC 1080 & 2009 July 15 & 48,48,45,44 & 55,56,60,61 \\
\enddata
\end{deluxetable}

\begin{deluxetable}{lcc|cc}
\tabletypesize{\scriptsize}
\tablewidth{0pt}
\tablecaption{Inferred sizes of Br$\gamma$ emission regions
\label{tab:brgsizes}}
\tablehead{\colhead{Source}
& \colhead{$\theta_{\rm Br \gamma}$ (mas)} &  \colhead{$R_{\rm
    Br\gamma}$ (AU)} & \colhead{$\theta_{\rm continuum}$ (mas)} &
\colhead{$R_{\rm cont}$ (AU) }}
\startdata
RY Tau & $<$0.12 & $<$0.01 & 2.63 $\pm$ 0.03 & 0.18 $\pm$ 0.01 \\ 
DG Tau & 1.85 $\pm$ 0.04 & 0.13 $\pm$ 0.01 & 2.44 $\pm$ 0.02 & 0.17 $\pm$ 0.01 \\ 
DK Tau & 1.26 $\pm$ 0.07 & 0.09 $\pm$ 0.01 & 1.63 $\pm$ 0.06 & 0.11 $\pm$ 0.01 \\ 
DR Tau & 0.68 $\pm$ 0.16 & 0.05 $\pm$ 0.01 & 1.72 $\pm$ 0.04 & 0.12 $\pm$ 0.01 \\ 
MWC 480 & $<$0.10 & $<$0.01 & 2.75 $\pm$ 0.01 & 0.19 $\pm$ 0.01 \\ 
RW Aur & $<$0.10 & $<$0.01 & 1.39 $\pm$ 0.07 & 0.10 $\pm$ 0.01 \\ 
MWC 758 & $<$0.21 & $<$0.02 & 2.36 $\pm$ 0.02 & 0.18 $\pm$ 0.01 \\ 
AS 205 A & $<$0.21 & $<$0.02 & 2.02 $\pm$ 0.07 & 0.16 $\pm$ 0.01 \\ 
MWC 863 & 0.97 $\pm$ 0.19 & 0.07 $\pm$ 0.01 & 3.84 $\pm$ 0.01 & 0.29 $\pm$ 0.01 \\ 
V2508 Oph & 0.56 $\pm$ 0.42 & 0.04 $\pm$ 0.03 & 3.81 $\pm$ 0.02 & 0.30 $\pm$ 0.01 \\ 
MWC 275 & 0.45 $\pm$ 0.40 & 0.03 $\pm$ 0.02 & 3.13 $\pm$ 0.02 & 0.19 $\pm$ 0.01 \\ 
AS 353 & 1.68 $\pm$ 0.05 & 0.13 $\pm$ 0.01 & 1.36 $\pm$ 0.06 & 0.10 $\pm$ 0.01 \\ 
V1057 Cyg & $<$5.00 & $<$1.50 & 1.27 $\pm$ 0.07 & 0.38 $\pm$ 0.02 \\ 
V1331 Cyg & 0.80 $\pm$ 0.13 & 0.28 $\pm$ 0.05 & 0.86 $\pm$ 0.12 & 0.30 $\pm$ 0.04 \\ 
MWC 1080 & 0.42 $\pm$ 0.44 & 0.21 $\pm$ 0.22 & 2.66 $\pm$ 0.03 & 1.33 $\pm$ 0.02 \\ 
\enddata
\tablecomments{Angular ring diameters ($\theta$) are
  converted into linear ring radii using the distances listed in Table
\ref{tab:sample}.}
\end{deluxetable}

\begin{deluxetable}{l|cccccc|ccccccccc}
\tabletypesize{\scriptsize}
\tablewidth{0pt}
\tablecaption{Results of Modeling
\label{tab:results}}
\tablehead{\colhead{Source}
& \colhead{$\chi_r^2$} & \colhead{$R_{\rm in}$} & \colhead{$PA$} & 
\colhead{$i$} & \colhead{$L/C$} & \colhead{$\alpha$} &
 \colhead{$\chi_r^2$} & \colhead{$R_{\rm in}$} & \colhead{$R_{\rm out}$} &
\colhead{$PA$} & \colhead{$\beta$} & \colhead{$\Delta v$} & \colhead{$L/C$} & 
\colhead{$f_{\rm a}$} & \colhead{$\alpha$} \\
 & & (AU) & ($^{\circ}$) & ($^{\circ}$) & & & & (AU) & (AU) & ($^{\circ}$) & & (km s$^{-1}$)}
\startdata
 & \multicolumn{6}{c}{Disk Models} & \multicolumn{8}{c}{Infall/Outflow Models} \\
\hline
RYTau & 0.09 & 0.03 & 45 & 70 & 0.1 &  4 & 0.08 & 0.03 & 0.10 & 90 &  1 & 350 & 0.1 & 1.0 &  1 \\
DGTau & 0.14 & 0.05 &  0 & 45 & 0.1 &  3 & 0.14 & 0.01 & 0.50 &  0 &  1 & 350 & 0.5 & 1.0 &  1 \\
DKTau & 0.27 & 0.05 & 45 & 70 & 0.1 &  2 & 0.26 & 0.03 & 0.10 & 45 &  1 & 500 & 0.1 & 0.1 &  1 \\
DRTau & 0.39 & 0.01 & 90 & 45 & 0.5 &  2 & 0.40 & 0.01 & 0.50 & 45 &  2 & 250 & 0.5 & 1.0 &  3 \\
MWC480 & 0.86 & 0.03 & 90 & 45 & 0.5 &  4 & 0.51 & 0.03 & 0.10 &  0 &  2 & 250 & 0.5 & 0.1 &  3 \\
RWAur & 0.59 & 0.01 & 45 &  5 & 0.1 &  4 & 0.62 & 0.01 & 0.10 &  0 &  2 & 250 & 0.1 & 1.0 &  1 \\
MWC758 & 0.20 & 0.03 &  0 & 70 & 0.1 &  4 & 0.20 & 0.01 & 0.05 & 45 &  2 & 250 & 0.1 & 1.0 &  3 \\
AS205A & 0.52 & 0.05 &  0 & 45 & 0.1 &  4 & 0.54 & 0.03 & 0.10 &  0 &  2 & 350 & 0.1 & 1.0 &  2 \\
MWC863 & 0.44 & 0.03 &  0 & 45 & 0.5 &  4 & 0.35 & 0.01 & 0.05 &  0 &  1 & 250 & 0.5 & 0.1 &  1 \\
V2508Oph & 0.72 & 0.05 & 45 &  5 & 0.5 &  2 & 0.76 & 0.01 & 0.50 &  0 &  2 & 250 & 0.5 & 1.0 &  1 \\
MWC275 & 0.54 & 0.01 &  0 & 45 & 0.5 &  4 & 0.30 & 0.01 & 0.05 & 90 &  2 & 250 & 0.5 & 0.5 &  3 \\
AS353 & 0.77 & 0.05 &  0 &  5 & 1.0 &  2 & 0.73 & 0.03 & 0.50 &  0 &  1 & 250 & 1.0 & 1.0 &  2 \\
V1057Cyg & 0.54 & 0.01 & 90 & 70 & 0.1 &  4 & 0.47 & 0.03 & 0.05 &  0 &  1 & 500 & 0.1 & 1.0 &  3 \\
V1331Cyg & 0.56 & 0.05 & 90 &  5 & 0.5 &  2 & 0.49 & 0.01 & 0.50 & 90 &  1 & 350 & 1.0 & 1.0 &  1 \\
MWC1080 & 0.28 & 0.03 & 90 & 45 & 0.5 &  2 & 0.26 & 0.03 & 0.10 & 45 &  2 & 500 & 0.5 & 1.0 &  1 \\
\enddata
\end{deluxetable}

\begin{deluxetable}{lccccccccc}
\tabletypesize{\scriptsize}
\tablewidth{0pt}
\tablecaption{Properties Derived from Br$\gamma$ Spectra
\label{tab:brg}}
\tablehead{\colhead{Source} & \colhead{EW (\AA)} & \colhead{$L_{\rm Br \gamma}$/($10^{-4}$ L$_{\odot}$)} & 
\colhead{$L_{\rm acc}$/L$_{\odot}$} & \colhead{$L_{\rm acc}$/L$_{\ast}$} & \colhead{$\dot{M}/(10^{-8} {\rm \: M_{\odot} \: yr^{-1}})$}}
\startdata
RYTau &    -1.0 &     1.9 &     0.5 &     0.1 &     6.4 \\
DGTau &    -8.1 &     3.8 &     1.3 &     1.5 &     7.6 \\
DKTau &    -2.2 &     0.8 &     0.2 &     0.2 &     1.2 \\
DRTau &    -7.1 &     3.1 &     1.0 &     1.1 &     5.7 \\
MWC480 &    -7.7 &     9.8 &     4.4 &     0.2 &     9.3 \\
RWAur &    -7.2 &     2.7 &     0.9 &     0.5 &     4.8 \\
MWC758 &    -1.4 &     1.9 &     0.6 &     0.1 &     1.1 \\
AS205A &    -2.6 &     3.7 &     1.3 &     1.0 &     8.1 \\
MWC863 &    -5.5 &     7.3 &     3.0 &     0.2 &     6.4 \\
V2508Oph &    -5.8 &     4.2 &     1.5 &     0.5 &    17.2 \\
MWC275 &    -6.1 &    14.7 &     7.3 &     0.2 &    17.5 \\
AS353 &   -12.6 &     1.5 &     0.4 &    &   \\
V1057Cyg &     0.9 &    &   &   &   \\
V1331Cyg &   -13.5 &    30.7 &    18.4 & &   \\
MWC1080 &    -5.1 &  1181.4 &  1824.8 &     0.1 &  4374.6 \\
\enddata
\end{deluxetable}

\epsscale{0.7}

\epsscale{0.8}

\epsscale{1.0}

\epsscale{0.8}

\epsscale{1.0}

\epsscale{0.8}
\clearpage

\begin{figure}
\plotone{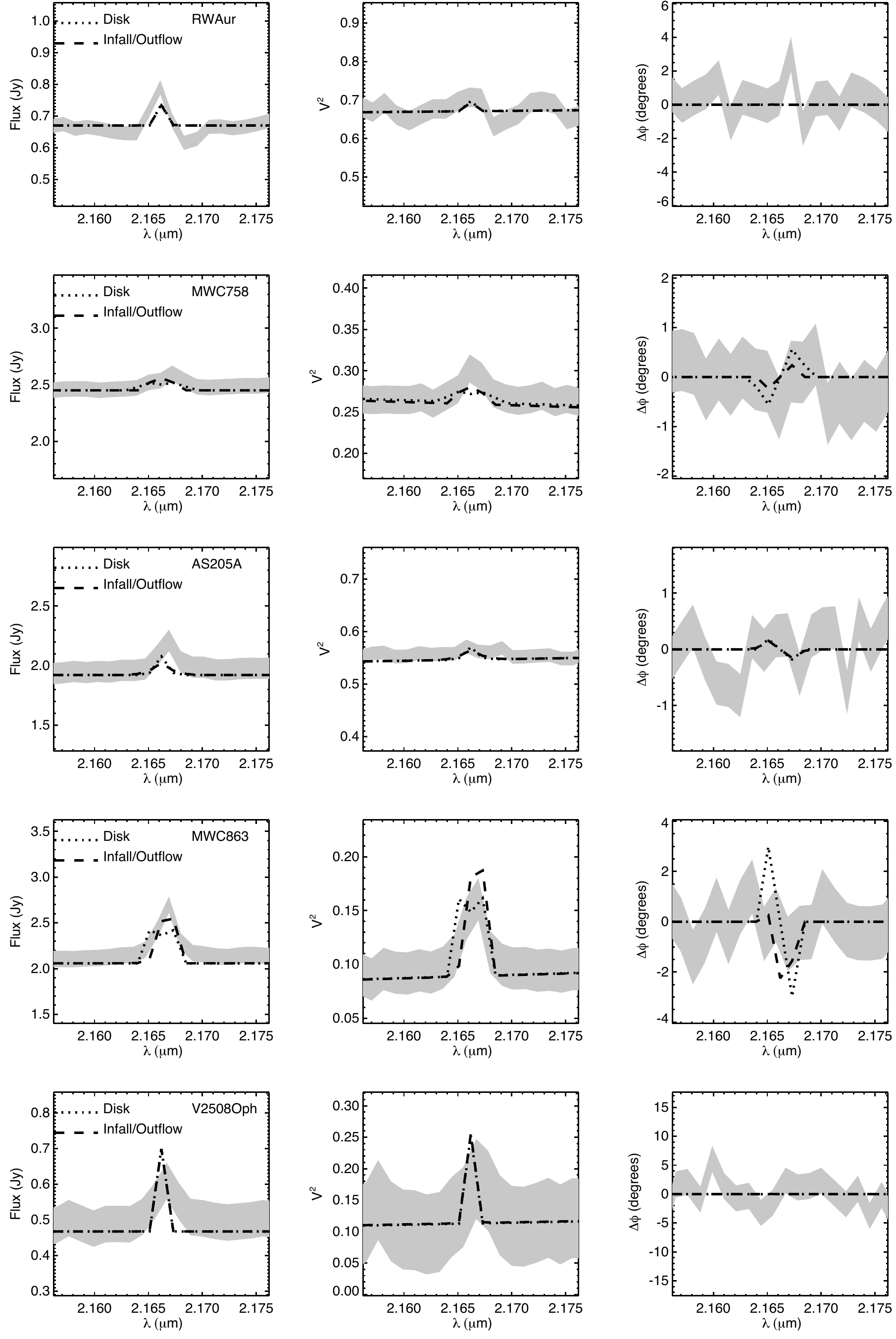}
\caption{Figure \ref{fig:modfits} continued.}
\end{figure}

\begin{figure}[tbhp]
\plotone{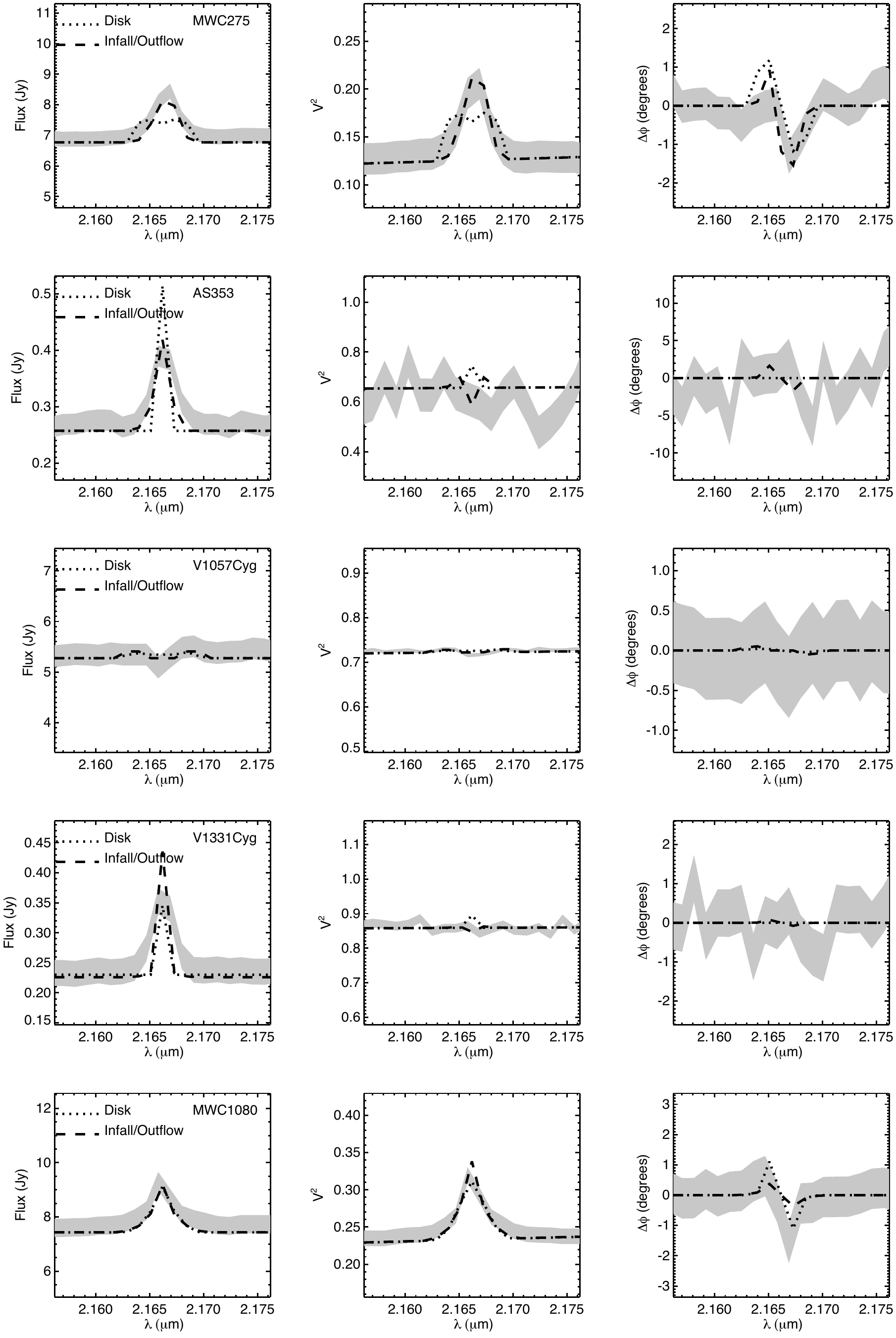}
\caption{Figure \ref{fig:modfits}  continued.}
\end{figure}

\epsscale{0.5}

%\epsscale{0.5}
%\begin{figure}[tbhp]
%\plotone{figs/Lacc_Bexp}
%\caption{Best-fit values of $\alpha$ for the infall/outflow models (from Table
%\ref{tab:results}) plotted against the accretion luminosities derived for these
%objects (from Table \ref{tab:brg}).
%\label{fig:lacc_alpha}}
%\end{figure}

\epsscale{0.5}

\end{document}